\documentclass[12pt]{iopart}
\usepackage[latin1]{inputenc}
\usepackage{amsmath}
\usepackage{amsfonts}
\usepackage{amssymb}
\usepackage{slashed}
\usepackage{graphicx}
\usepackage{url}
\usepackage{hyperref}
\usepackage{bookmark}

\newcommand{\bvec}[1]{\ensuremath{\boldsymbol{#1}}}
\newcommand{\dd}{\mathrm{d}}

\begin{document}
\topical{Neutrino event generators:\\ foundation, status and future}

\author{Ulrich Mosel}

\address{Institut fuer Theoretische Physik, Universitaet Giessen, Giessen, Germany}
\ead{mosel@physik.uni-giessen.de}
\vspace{10pt}
\begin{indented}
	\item[]\today
\end{indented}

\begin{abstract}

Neutrino event generators are an essential tool needed for the extraction of neutrino mixing parameters, the mass hierarchy and a CP violating phase from long-baseline experiments. In this article I first describe the theoretical basis and the approximations needed to get to any of the generators. I also discuss the strengths and limitations of theoretical models used to describe semi-inclusive neutrino-nucleus reactions. I then confront present day's generators with this theoretical basis by detailed discussions of the various reaction processes.  Finally, as examples, I then show for various experiments results of the generator GiBUU for lepton semi-inclusive cross sections as well as particle spectra. I also discuss features of these cross sections in terms of the various reaction components, with predictions for DUNE.  Finally, I argue for the need for a new neutrino generator that respects our present-day knowledge of both nuclear theory and nuclear reactions and is as much state-of-the-art as the experimental equipment. I outline some necessary requirements for such a new generator.

\end{abstract}

%\submitto{\JPG}

%
% For two-column output uncomment the next line and choose [10pt] rather than %[12pt] in the \documentclass declaration

\section*{Contents}
\tableofcontents
\title[Neutrino event generators]{}

\newpage

\section{Introduction}
Electron scattering on nuclei has been an active field of nuclear physics research over many decades. It has increased our knowledge about the response of a nuclear many-body system to the electromagnetic interaction \cite{Boffi:1993gs}. At relatively low energies (10s of MeV) collective excitations of the nucleus are dominant, at higher energies (100 MeV) quasielastic reactions on individual nucleons become essential \cite{Benhar:2006wy}, at still higher energies (100s of MeV) one enters the regime of nucleon resonance excitations and finally, at the highest energies (10s of GeV), the reactions explore the Deep Inelastic Scattering (DIS) regime \cite{Bianchi:2007fz}. Originally unexpected phenomena such as 2p2h excitations \cite{Dekker:1991ph}, in-medium spectral functions \cite{Benhar:1994hw,CiofidegliAtti:1995qe}, short-range correlations \cite{Alvioli:2008as}  and spectroscopic factors \cite{Radici:2002ut} and the change of parton distributions inside the nucleus, as showing up in the EMC effect \cite{Hen:2016kwk}, have been explored. For all of these features not only the incoming beam energy is important, but in addition also the momentum transfer.

It is, therefore, natural to extend these studies to reactions with neutrinos where the axial coupling, typical for weak interactions, offers a new degree of freedom to explore \cite{Bernard:2001rs}. Indeed, such studies were originally motivated by the interest in the axial response of nuclear many-body systems. In the first such studies, nearly 50 years ago, the nucleus was described as an unbound system of freely moving nucleons with their momenta determined by the Fermi-gas distribution \cite{Smith:1972xh,LlewellynSmith:1971zm}. First steps beyond that simple model were studies where the nucleus was described as a bound system with a mean-field potential \cite{Donnelly:1984yr,Alberico:1997vh,Amaro:2011qb,Meucci:2015bea} and excitations were treated by the Random Phase Approximation (RPA)\cite{Kolbe:1995af,Nieves:2004wx,Botrugno:2005kn,Pandey:2014tza}. More recently, even ab-initio calculations of the electroweak response of nuclei have become possible \cite{Lovato:2017cux}. In general, neutrino-induced reactions exhibit the same characteristics as the electron-induced reactions. The same reaction subprocesses as described above also are present here. The only difference being the additional presence of an axial amplitude in all processes; the final state interactions are the same \cite{Mosel:2016cwa} if the incoming kinematical conditions (energy- and momentum-transfer) are identical.

Even though these two types of experiments, electron-induced and neutrino-induced ones, are so similar there is a very essential differences between them. In electron-induced reactions the incoming beam energy is very accurately known and the momentum transfer can be measured with the help of magnetic spectrometers. Both of these observables are not available for neutrinos. Because neutrino beams are produced through the secondary decay of pion and kaons, first produced in a p+A reaction, their energies are not sharp, but smeared out over a wide range. For example, for the Deep Underground Neutrino Experiment (DUNE) \cite{DUNE} the energy distributions peaks at about 2 GeV, but has long tails all the way down to zero energy and up to 30 GeV. In a charged current (CC) reaction the outgoing lepton's energy and angle can be measured, but since the incoming energy is not known, also the momentum transfer is experimentally not available. This is per se a challenge for any comparison of theory with experimental results since the theory calculations have to be performed at many different energies. It also presents a challenge to theory because even lepton semi-inclusive cross sections cannot be easily separated according to the first interaction process since the smearing of the incoming energy automatically brings with it a smearing of the energy- and momentum-transfer. For example, even at the lower energies of the T2K experiment (beam energy peak at about 0.75 GeV \cite{t2k}) true quasielastic (QE) scattering on a single nucleon cannot be separated from events involving 2p2h interactions or those in which first a pion was created that was subsequently reabsorbed. Thus, any theoretical description of experimental data requires a simultaneous and consistent treatment of several different elementary interaction processes.

Presently running (T2K \cite{t2k}, NOvA \cite{nova}) or planned (T2HK\cite{T2HK}, DUNE\cite{DUNE}) oscillation experiments aim at a precise determination of the neutrino mixing parameters, of a CP violating phase and of the mass ordering of neutrinos; they all use nuclei as targets. The oscillation formulas used to extract these quantities from the data all involve the incoming neutrino energy. This incoming energy must be reconstructed from the measured final state of the neutrino-nucleus reaction. Because of experimental acceptance cuts and the entanglement of different elementary processes this reconstruction is less than trivial. It involves an often wide extrapolation from the actually measured final state to the full final state.

\paragraph{Energy reconstruction}
There are two methods in use to reconstruct the energy of the incoming neutrino from final state properties:
\begin{itemize}
\item {\bf Kinematical Method} In this method one uses the fact that the incoming energy of a neutrino interacting in a CCQE process with a neutron at rest is entirely determined by the kinematics of the outgoing lepton. If this method is used with nuclear targets then there are two complications:
  \begin{itemize}
  	\item First, the neutron is not free and at rest, but it is bound in a nucleus and Fermi-moving. This alone already leads to a smearing of the reconstructed energy and a shift with respect to the free-nucleon case.
  	\item Also, the method in principle requires to identify the interaction process as true QE scattering. 
  	   \begin{itemize}
  	    \item This, however, is impossible in a nuclear target where always pion production followed by reabsorption can take place. This effect leads to a low-energy tail in the reconstructed energy \cite{Leitner:2010kp}. Therefore, the method is expected to work best at lower energies, where pion production is not yet dominant. 
  	    \item In addition, detector acceptances may limit the necessary QE identification.
  	   \end{itemize} 
  \end{itemize}	

\item {\bf Calorimetric Method} In this method one measures the energies of all particles in the final state and reconstructs the energy from that information. Problems with this method arise because detectors are not perfect, have detection threshold and may miss certain particles, e.g. neutrons, alltogether. The energy then has to be reconstructed from a final state that is only partially known. This method is mostly used in higher-energy experiments where the complication due to inelastic excitations and pion production make the kinematical method less reliable.

\end{itemize}

\begin{figure}
\centering
\includegraphics[width=0.7\linewidth]{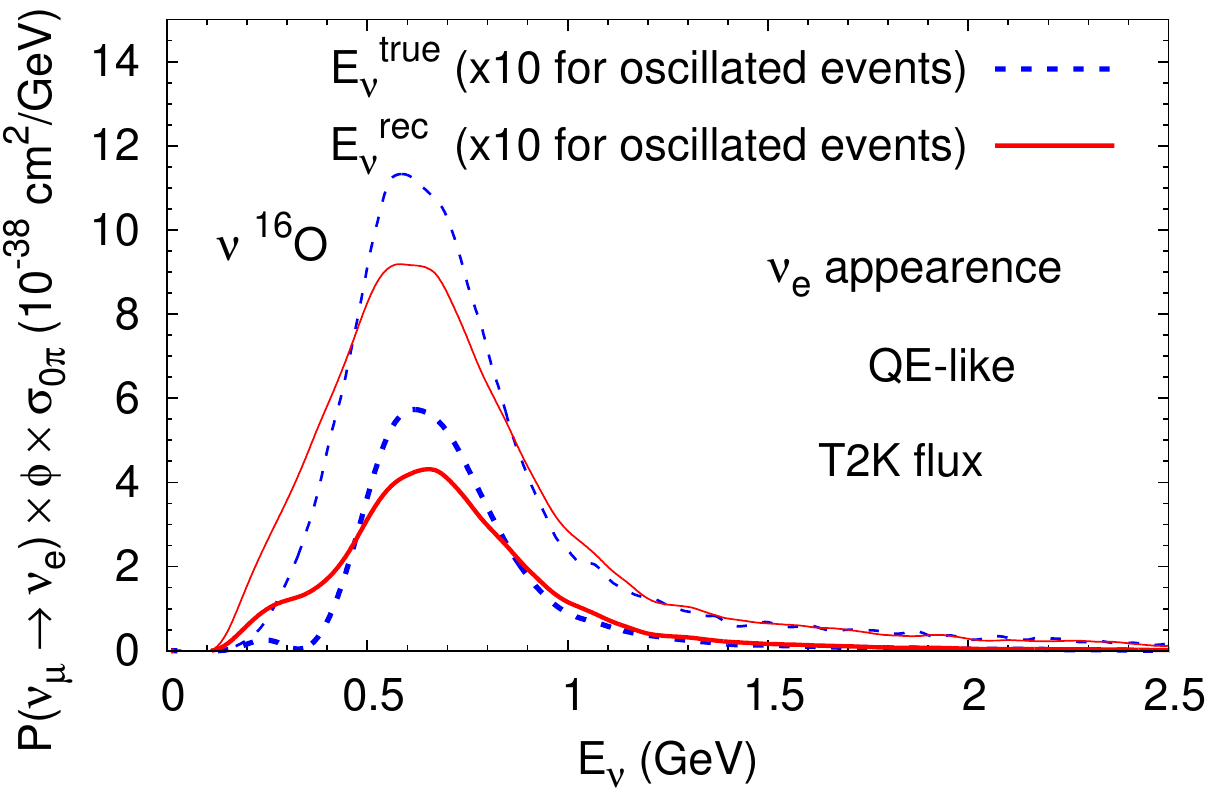}
%\end{figure}
%\begin{figure}
%\centering
\includegraphics[width=0.7\linewidth]{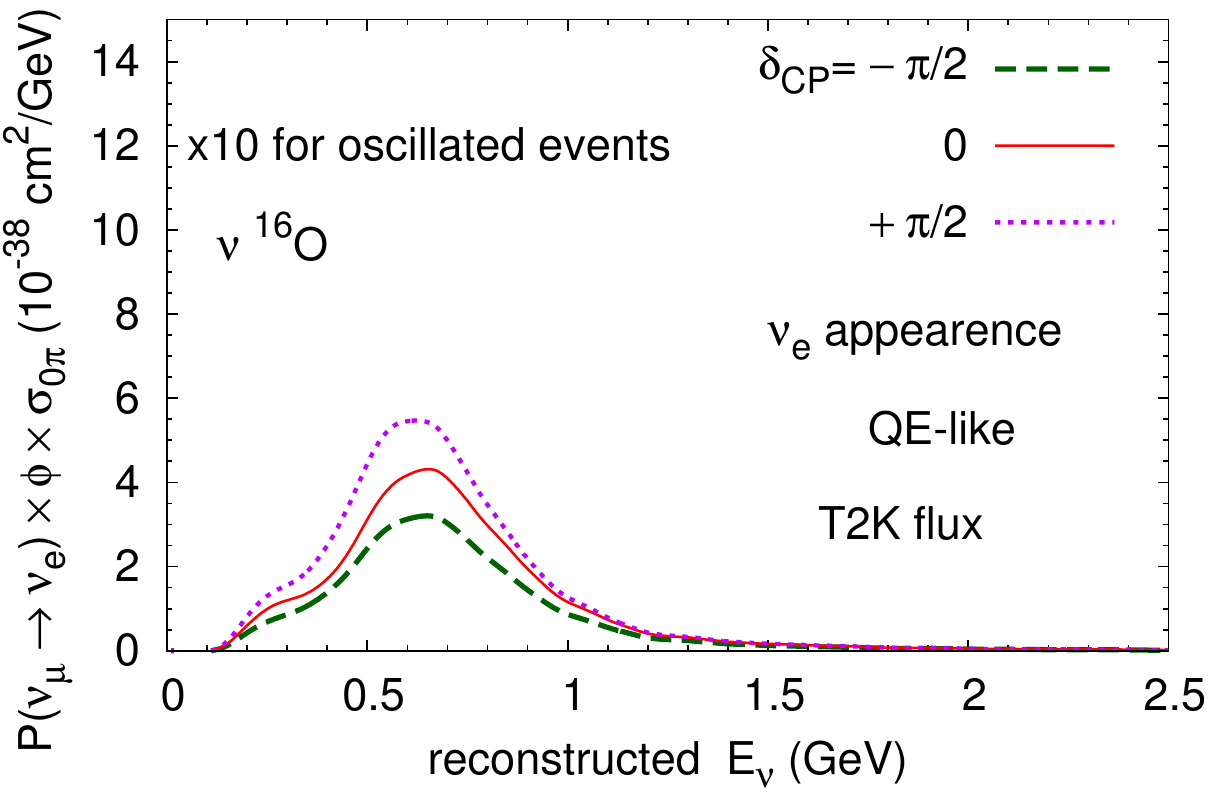}
\caption{{\bf Top:} QE-like event distributions in the T2K flux for electron neutrino appearance experiments, as obtained from GiBUU. The dashed curves give the event distributions as function of the true incoming neutrino energy, the solid curves those as a function of reconstructed energies. The oscillated event curves have been multiplied by a factor of 10 to enhance the visibility of the difference.
{\bf Bottom:} Sensitivity of QE-like event distributions on the CP violating phase. The solid (red) curve is the same as the one in the upper part. Both figures taken from \cite{Lalakulich:2012hs}.}
%\label{fig:t2kflux-oscillations-deltacp}
\label{fig:t2kflux-oscillations}
\end{figure}

An illustrative example for the typical errors in energy reconstruction is shown in the upper part of Figure \ref{fig:t2kflux-oscillations}. For this example the kinematical method was used.
Here a generator, in this case GiBUU \cite{gibuu,Buss:2011mx}, has been used to generate millions of events as a function of 'true' energy. These events were then analyzed and the energy was reconstructed by using the so-called kinematical method. The peak of the oscillated event distribution lies at about 0.65 GeV; here the true and the reconstructed curves differ by about 25\%. This discrepancy is just as large as the sensitivity of the electron appearance signal to the CP violating phase $\delta_{CP}$. This is illustrated for exactly the same reaction (T2K flux on $^{16}$O) in the lower part of Figure \ref{fig:t2kflux-oscillations}.
At the peak of the oscillation signal the curves corresponding to three different values of $\delta_{CP}$ differ by about 25\%, i.e.\ just by about the same amount as the error in the energy reconstruction. The influence of this error on neutrino mixing parameters has been quantified in Ref.\ \cite{Coloma:2013rqa,Ankowski:2015jya}.

In both methods for the energy reconstruction the actually measured energy has to be extrapolated to the true one. To perform this extrapolation and reconstruction so-called neutrino generators have been constructed early on. These generators try to take not only the initial neutrino-nucleon interaction into account, but also the quite essential final state interactions. A good review of generators presently used by experiments is given in \cite{Gallagher:2018pdg}. There it is also pointed out that generators are also needed for simulations of practical importance, e.g.\ for acceptance studies and for handling the effects of the typically quite large, extended targets. While the generator NEUT \cite{Hayato:2002sd} is primarily being used by the T2K experiment, the generator GENIE \cite{GENIE,Andreopoulos:2009rq} has become widely used by groups connected to Fermilab experiments, such as MicroBooNE, NOvA and MINERvA. In addition, a generator named NuWro \cite{Golan:2012rfa} is being used for comparisons of experiment with calculations. Also a transport theoretical framework, GiBUU \cite{gibuu,Buss:2011mx}, for general nuclear reactions can be used as a generator of neutrino interactions with nuclei.

\paragraph{Review outline}

Obviously, in all these generators cross sections both for the initial neutrino-nucleon reaction and for the hadron-hadron reactions in the final state are essential \cite{Katori:2016yel}. Unfortunately, the few data that exist on elementary targets, such as p and D, are all at least 30 years old and carry large uncertainties. Our knowledge about these cross sections has been discussed in a number of fairly recent reviews \cite{Conrad:1997ne,Gallagher:2011zza,Formaggio:2013kya,Mahn:2018mai}.
In this article I will, therefore, not repeat the discussions of cross sections.

Instead I first give a short outline of the theoretical basis for {\it any} generator and discuss the approximations that go into the presently used ones. I will then go through the various subprocesses (QE, Resonance excitation, DIS, ...) and confront the inner workings of the generators with present-day nuclear physics knowledge about these processes. The main motivation for this critical discussion is my conviction that only a theoretically up-to-date and consistent generator can provide the reliability needed when used for new targets or in new energy regimes.

While all of these discussions are generally valid I will then illustrate features of neutrino-nucleus cross sections with the help of a specific generator, GiBUU. Towards the end of this critical review I will argue that in view of the upcoming high-precision experiments also a new well-founded, high-precision generator is needed that is free of many the shortcomings of presently used ones.

\section{Foundations of generators} \label{Generators}
In this chapter I briefly discuss the theoretical basis of \emph{all} generators. Most of the generators treat the hadrons as billiard balls following classical trajectories. It is, therefore, essential to understand under which circumstances such a treatment can be justified. In the following subsection I merely summarize the essential steps necessary to get to a well-founded transport equation; more details can be found in \cite{Buss:2011mx,Kad-Baym:1962,Danielewicz:1982kk,Danielewicz:1982ca}.

\subsection{Short derivation of a general transport equation}
The dynamical development of any quantum mechanical many-body system is determined by an infinite set of coupled equations for the Green's functions: the one-particle Green's function depends on the two-particle one, the two-particle Green's function depends on the higher-order one, and so on. All the higher order Green's functions can formally be included in a self-energy. The dynamics of the correlated many-body system is then determined by a Dyson equation for the single particle Green's function which contains the (quite complicated) self-energy. In addition, interaction vertices of single particles can be dressed.

Now approximations are introduced:
\begin{itemize}
	
\item The first, and most important, approximation is to truncate the hierarchy of coupled Green's functions by neglecting all higher-order correlations and keeping only the single-particle Green's function. Better is the so-called Dirac-Brueckner-Hartree-Fock (DBHF) approach in which some two-body correlations are being taken into account. The self-energies are modeled, e.g.\ by an energy-density functional theory or the relativistic mean field theory.

\item It is, furthermore, assumed that all particles move locally in a homogeneous medium, with corresponding self-energies (potentials). This is the so-called local-density approximation.

\item The damping terms, i.e.\ the widths, of all the particles in medium are small relative to the mass gap. In nuclear physics this is well fulfilled since spectral functions of nucleons inside nuclei are very narrow compared to the total mass \cite{Benhar:1992cnb}.

\end{itemize}
Closely connected with the local-density approximation is the 'gradient-approximation' in which it is assumed that the Green's functions $G(x,x')$ are rapidly oscillating functions of the relative coordinates $x - x'$ while their variation with the center of mass coordinate $X = (x + x')/2$ is small\footnote{Here $x$ is the space-time four-vector}. This is the case if the medium itself is nearly homogeneous. For a nuclear system it implies that this approximation is the better the heavier the nucleus is; in heavy nuclei the homogeneous volume part prevails over the inhomogeneous surface region.

For a lepton-nucleus reaction the starting point are the single particle Green's functions for the nucleons in the target and the incoming lepton. In a homogeneous system it is advantageous to introduce the so-called Wigner-transforms of a single particle Green's function
\begin{eqnarray}
G^<_{\alpha \beta}(x,p) &=& \int d^4\xi\, {\rm e}^{i p_\mu\xi^\mu}\, ({\rm i}) \left< \bar \psi_\beta(x+\xi/2) \psi_\alpha (x-\xi/2) \right>  \nonumber \\
G^>_{\alpha \beta}(x,p) &=& \int d^4\xi\, {\rm e}^{i p_\mu\xi^\mu}\, ({\rm -i}) \left< \psi_\alpha(x-\xi/2) \bar \psi_\beta (x+\xi/2) \right>
\end{eqnarray}
which depend on the Dirac indices. These are the objects that determine the time-evolution in a lepton-nucleus reaction. They are nothing else than the relativistic one-body density matrices.  By tracing over the Dirac indices (i.e.\ concentrating on the spin-averaged behavior) one obtains a vector current density
\begin{equation}   \label{FV}
F^\mu_V(x,p)= -{\rm i\, tr}\left(G^<(x,p)\gamma^\mu\right) ~.
\end{equation}

The time development of the vector current density for Dirac particles, i.e.\ leptons and nucleons, is then given by
\begin{eqnarray}     \label{transp}
\partial_\mu F_V^\mu(x,p) &-& {\rm tr}\left[\Re \Sigma^{\rm ret}(x,p), -{\rm i} G^<(x,p)\right]_{\rm PB} \nonumber \\
&+& {\rm tr} \left[\Re G^{\rm ret}(x,p),-{\rm i} \Sigma^<(x,p)\right]_{\rm PB} = C(x,p) ~.
\end{eqnarray}
with
\begin{equation}
C(x,p) = {\rm tr} \left[\Sigma^<(x,p)G^>(x,p) - \Sigma^> (x,p)G^<(s,p)\right] ~.
\end{equation}
In Eq.\ (\ref{transp}) the symbol $\left[\dots\right]_{\rm PB}$ stands for the Poisson bracket
\begin{equation}
\left[S,G\right]_{\rm PB} = \frac{\partial S}{\partial p_\mu} \frac{\partial G}{\partial x^\mu} -  \frac{\partial S}{\partial x^\mu} \frac{\partial G}{\partial p_\mu}
\end{equation}
and the quantities $\Sigma$ in (\ref{transp}) are self-energies which represent
the potentials and are thus essential ingredients of the Hamiltonian.

In a homogeneous system of fermions one can relate the two propagators $G^>$ amd $G^<$ to each other
\begin{eqnarray} \label{G><}
{\rm i} G^<(x,p) =  + 2 f(x,p)\, \Im G^{\rm ret}(x,p) \nonumber \\
{\rm i} G^>(x,p) = -2 (1 - f(x,p)) \,\Im G^{\rm ret}(x,p)   ~,
\end{eqnarray}
where $f(x,p)$ is a Lorentz-scalar function. The trace over the imaginary part of the retarded propagator in (\ref{G><}) is -- up to some numerical factors -- just the single particle spectral function $A(x,p)$. Reducing the vector current density $F_V^\mu$ to a scalar density $F$ by means of $F_V^\mu = ({p^*}^\mu/E^*) F$, where $p^*$ and $E^*$ are the momentum and the energy for a particle with self-energies, one has from (\ref{FV})
\begin{equation}   \label{def_F}
F(x,p) = 2 \pi g f(x,p) A(x,p) ~.
\end{equation}
The function $F$ is the actual density distribution function in the eight-dimensional phase space $(x,p)$. It thus describes the time-development also of off-shell particles. Since it contains also the spectral function, often $F$ is called the 'spectral phase space density' whereas $f(x,p)$ is just the 'phase space density'. The factor $g$ is a spin-degeneracy factor.

The equation of motion for $F_V$ can now be converted into one for $F$. It becomes
\begin{equation}     \label{BUU}
\mathcal{D}F(x,p) + {\rm tr}\, \left[\Re G^{\rm ret}(x,p), - {\rm i} \Sigma^<(x,p)\right]_{\rm PB} = C(x,p)
\end{equation}
with
\begin{equation}   \label{drift}
\mathcal{D}F = \left[p_0 - H,F \right]_{\rm PB}
\end{equation}
 This so-called 'drift term' (\ref{drift}) originates in the first Poisson bracket of Eq.\ (\ref{transp}). $H$ is the single particle Hamiltonian which involves Lorentz-scalar and -vector potential mean fields (including in particular the Coulomb field); it is obtained from Eq. (\ref{transp}) by identifying the selfenergy given there by a potential.

The physics content of the second term on the left hand side of Eq.\ (\ref{BUU}) is not obvious. It is also not easy to handle numerically since the  term does not explicitly contain the spectral phase space density $F$. A major simplification was achieved by Botermans and Malfliet \cite{Botermans:1990qi} who showed that this term can be evaluated under the assumptions of local equilibrium in phase space and the gradient approximation. The equation of motion for $F$ then becomes
\begin{equation}   \label{GiBUU}
\mathcal{D}F(x,p) - {\rm tr} \Big\{ \Gamma(x,p)f(x,p),\Re G^{\rm ret}(x,p) \Big\}_{\rm PB} = C(x,p)   ~.
\end{equation}
Now the second term on the lhs is proportional to $F$ thus simplifying its practical evaluation. The quantity $\Gamma(x,p)$ is the imaginary part (width) of the retarded self-energy. This shows that this term is connected to the in-medium width and is essential for off-shell transport: its presence ensures that, e.g., the in-medium spectral function of a nucleon  becomes a $\delta$-function when the nucleon leaves the nucleus.

Using Eqs.\ (\ref{G><}) and (\ref{def_F}) the term $C(x,p)$ on the rhs of Eq.\ (\ref{BUU}) becomes
\begin{equation}
C(x,p) = 2 \pi g \, {\rm tr} \Big\{ \left[ \Sigma^>(x,p) f(x,p) -  \Sigma^<(x,p) (1 - f(x,p) ) \right] \, A(x,p) \Big\} ~.
\end{equation}
This term has the typical structure of a loss term (1. term in parentheses) that is proportional to the phase-space density of the interacting particle and a gain term (2. term) that takes the Pauli-principle into account. The self-energies $\Sigma^{\stackrel{>}{<}}$ contain the transition probabilities for both processes. $C(x,p)$ thus represents a collision term that that takes into account that interactions with other particles can either deplete a specific phase-space volume or populate it; in the latter case the Pauli-principle (for fermions) is taken into account by the factors $(1 - f)$.

Eq.\ (\ref{BUU}) without the second (off-shell transport) term has the form of the Boltzmann-Uehling-Uhlenbeck (BUU) equation.

Eq.\ (\ref{GiBUU}) represents the center piece of a practical off-shell transport theory; it is, e.g., encoded in the generator GiBUU \cite{gibuu,Buss:2011mx}. For each particle there is one such equation to be solved and they are all coupled through the collision term, on one hand, and by the mean field potentials in $H$ (to which all particles contribute), on the other hand. If particles, such as e.g.\ pions, are created in a collision the corresponding equation for them has to be added to the initial system of equations \footnote{For bosons, e.g.\ pions, the equation actually looks slightly different, see \cite{Buss:2011mx}}. For an explicit example consider the case of a neutrino-nucleus interaction: initially then there is one such equation for the incoming neutrino with a $\delta$-function like spectral function ($\Gamma=0$) and {\it A} equations which contain the spectral phase-space densities of the {\it A} nucleons in the nuclear ground state; their spectral functions are contained in $F$ and in the Botermans-Malfliet off-shell transport term.

At this point it is worthwhile to point out that the theory developed so far as expressed in Eq.\ (\ref{BUU}) is fully relativistic and the equations of motion are covariant.

\subsubsection{Initial conditions}
The initial conditions for the integrations of the transport equation of motion (\ref{GiBUU}) are determined by the spectral phase-space density at time $t=0$
\begin{equation}   \label{spectrphsp}
F(x,p)_{t=0} = 2 \pi g f(\bvec{x},0,p) A(\bvec{x},0,p) ~.
\end{equation}
and is thus fully determined by the Wigner-transform of the one-body density matrix. This density matrix could be obtained from any nuclear many-body theory.

\subsection{Numerical methods}
The generalized BUU equation (\ref{GiBUU}) can be solved numerically by using the
test-particle technique, i.e., the continuous Wigner function is
replaced by an ensemble of test particles represented by
$\delta$-functions,
\begin{equation}
  F(x,p)= \lim_{n(t)\to \infty}\frac{{\left( 2\pi \right) }^4}{N}
  \sum_{j=1}^{n(t)} \delta[\bvec{x}-\bvec{x}_j(t)]
  \delta[\bvec{p}-\bvec{p}_j(t)] \delta[p^0-p^0_j(t)]~,
  \label{eq:testparticleansatz}
\end{equation}
where $n(t)$ denotes the number of test particles at time, $t$, and
$\bvec{x}_j(t)$ and $p_j(t)$ are the coordinates and the four-momenta of
test particle $j$ at time $t$. As the phase-space density changes in
time due to both, collisions and the mean field dynamics, also the number of
test particles changes throughout the simulation: in the collision term,
test particles are destroyed and new ones are created, for example when a pion
is absorbed or produced. At $t=0$ one starts
with $n(0)=N \cdot A$ test particles, where $A$ is the number of
physical particles and $N$ is the number of ensembles (test particles
per physical particle). More details about the numerical treatment of
the Vlasov and collision dynamics can be found in \cite{Buss:2011mx}.

While this method is well established for the drift term of the BUU equation the collision term requires some more refinement.
Here one has often just used a geometrical argument to relate a cross section between two particles $\sigma = \pi d^2$ to an interaction distance
$d$. This recipe poses a problem when the energies of the interacting particles become relativistic since then the distance
seen from either one of the two particles may be different in their respective restframes because there are two different eigentimes for the two particles involved whereas the equation itself contains only the laboratory time. However, in the context of heavy-ion collisions approximate schemes have been developed that minimize this problem; these same methods can also be used in neutrino generators. Other schemes involve interactions
between all particles in a given phase-space cell \cite{Lang:1993}; any relativistic deficiencies can be minimized that way. The actual
choice of a particular reaction channel is then done by a cross section weighted random decision.

\subsection{Approximations}

\subsubsection{Quasiparticle approximation}
In the quasiparticle approximation one neglects the width of the single particle spectral function of all particles. This gives
\begin{equation}  \label{QPapprox}
F(x,p) = 2 \pi g \, \delta[p_0 - E(x,\mathbf{p})]\, f(x,\mathbf{p})
\end{equation}
Here $E(x,\mathbf{p})$ is the energy of a particle in mean field that depend on $x$ and $\mathbf{p}$ and $g$ is the spin-isospin degeneracy. In this approximation the off-shell transport term in (\ref{GiBUU}) disappears and the equation becomes for a $2 \rightarrow 2'$ collision
\begin{equation}   \label{QPBUU}
\begin{split}
\Big[ \partial_t + & (\bvec{\nabla}_{\bvec{p}} E_{\bvec{p}}) \cdot
\bvec{\nabla}_{\bvec{x}} - (\bvec{\nabla}_{\bvec{x}} E_{\bvec{p}})
\cdot \bvec{\nabla}_{\bvec{p}} \Big ] f(x,\bvec{p}) = \frac{g}{2}
\int \frac{ \dd^3 \bvec{p}_2 \; \dd^3 \bvec{p}_1' \; \dd^3
	\bvec{p}_2'}{(2 \pi)^9} \frac{m_{\bvec{p}}^* m_{\bvec{p}_2}^*
	m_{\bvec{p}_1'}^*
	m_{\bvec{p}_2'}^*}{E_{\bvec{p}}^* E_{\bvec{p}_2}^* E_{\bvec{p}_1'}^* E_{\bvec{p}_2'}^*} \\
& \times (2 \pi)^4 \delta^{(3)}(\bvec{p}+\bvec{p}_2 - \bvec{p}_1' -
\bvec{p}_2') \delta(E_{\bvec{p}}+E_{\bvec{p}_2}-E_{\bvec{p}_1'}-E_{\bvec{p}_2}') \times  \overline{|\mathfrak{M}_{p\,p_2 \to p_1'\,p_2'}|^2}\\
& \times [ f(x,\bvec{p}_1') f(x,\bvec{p}_2') \overline{f}(x,\bvec{p}) \overline{f}(x,\bvec{p}_2)
- f(x,\bvec{p}) f(x,\bvec{p}_2) \overline{f}(x,\bvec{p}_1') \overline{f}(x,\bvec{p}_2')]
\end{split}
\end{equation}
with $E_p=E(x,\mathbf{p})$. The functions $\bar f = 1 - f$ contain the effects of the Pauli-principle.
The transition probability averaged over spins of initial particles and summed over spins of final particles is denoted by $\overline{|\mathfrak{M}_{p\,p_2 \to p_1'\,p_2'}|^2}$. It has to be calculated with final states that contain the effects of the same potential as the one in the drift term. The stars "*" denote in-medium masses and energies that involve potentials. The corresponding expressions for other collisions, such as $2 \rightarrow 3$, or the decay of a resonance $1 \rightarrow 2 + 3$ can be found in \cite{Buss:2011mx}.

The quasiparticle approximation describes a system of particles that move in a potential well. This is thus obviously a reasonable description not only of a nuclear groundstate, but also the final state interactions that take place in this same potential. The phase-space distributions $f(x,\mathbf{p})$ are the same in the drift term as in the collision term. If many different reaction channels are open, e.g.\ at T2K energies CCQE scattering and $\Delta$ resonance excitation, the collision term consists of a sum of terms for the various reaction processes. Essential is that for each individual reaction channel the initial ground state distribution $f$ of the nucleons is the same.

In the quasiparticle approximation one neglects the in-medium spectral function of particles. For nucleons this essentially amounts to neglecting their short-range correlations that are known to lead to a broadening of the nucleon's spectral function. For in vacuum unstable particles, which have already a free width (e.g. the $\Delta$ resonance), one has two possibilities: one can either treat these particles only as intermediate, virtual excitations that contribute to the transition matrix elements in the collision term, but are never explicitly propagated. For very broad, i.e.\ very short-lived, resonances this is a reasonable assumption. On the other hand, a resonance such as the $\Delta$ lives long enough to be propagated as an actual particle. This propagation could then be handled by propagating $\Delta$s with different masses, but it requires some knowledge about $\Delta N$ interactions in the collision term.

All neutrino generators so far  work in the quasiparticle approximation, although GiBUU allows also for off-shell propagation.

\subsubsection{Frozen approximation}
The mean field potentials contained in the Hamiltonian $H$ in (\ref{QPBUU}) depend self-consistently (in a Hartree-Fock sense) on the phase-space distributions of the target nucleons. If nucleons are being knocked out by the incoming neutrino then also the mean field changes. To take care of this time-dependent change of the target structure requires some numerical expense.

A significant computational simplification can be reached by assuming that the interaction is not violent enough to disrupt the whole nucleus, but allows just for the emission of a few ($\approxeq 2 -  3$) nucleons for nuclei with a mass number $A >12$. In this case it is reasonable to assume that the nuclear density distribution does not change significantly with time ("frozen approximation", sometimes also called "perturbative particle method"). Since at the same time the number of ejected particles is relatively small collisions take place only between already ejected particles and the frozen target nucleons, but not between ejected particles. This approximation obviously becomes the better the heavier the target nucleus is and the lower the incoming neutrino energy.

The frozen approximation is used in all the generators, but its actual implementation is quite different. Whereas standard generators freeze the density and then decide about final state interactions by means of a mean free path in that density, in GiBUU collisions between outgoing hadrons and the target nucleons are handled by picking out a target nucleon with its binding energy and its momentum in the Fermi sea. In this case the Fermi sea occupation is not changed during the time-development of the reaction.

\subsubsection{Free particle approximation}    \label{sect:QPapprox}
 Both GiBUU and FLUKA \cite{Battistoni:2009zzb} have potentials for the nucleons implemented so that the nuclear ground states are actually bound and some of the nucleon-nucleon effects are already incorporated in the potential. As a consequence the effects of residual interactions (e.g.\ RPA) are diminished.  A prize one has to pay for the presence of potentials is in terms of computer time. In between collisions the nucleons move on trajectories that are determined by the potentials; these trajectories have to be numerically integrated, if potentials (including Coulomb) are present.

The widely used neutrino generators GENIE and NEUT (as well as NuWro) do not contain any binding potentials. In these generators the nuclear ground state is not bound and the system of nucleons, initialized with a momentum distribution of either the global or the local Fermi gas model, would fly apart if the nucleons were propagated from time $t=0$ on. Target nucleons and ejected nucleons are thus being treated on a very different basis. While the former are essentially frozen, the latter are being propagated from collision to collision in the final state interaction phase. The phase-space distributions of all nucleons are always those of free nucleons with free dispersion relations connecting energy and momentum. This makes it numerically simple to follow the nucleons in the final state because in between collisions they move on straight-line trajectories. Binding energy effects are at the end introduced by fitting an overall binding energy parameter to final state energy distributions.

In the free particle approximation the structure of equation (\ref{QPBUU}) is quite transparent. For free on-shell particles one has \cite{Buss:2011mx}
\begin{eqnarray}   \label{QPeq}
H &=& p^2/(2M) \nonumber \\
F(x,p) &=& 2\pi g \delta(p_0 - E) f(x,\mathbf{p})
\end{eqnarray}
where $g$ is a spin-isospin degeneracy factor. Inserting this into (\ref{QPBUU}) gives
\begin{equation}   \label{Gentransp}
\left(\partial_t + \frac{\mathbf{p}}{M} \cdot \mathbf{\nabla_x}\right) f(x,\mathbf{p}) = C(x,\mathbf{p}) ~.
\end{equation}
In this equation all potentials and spectral functions are neglected, except for the collision term it describes the free motion of particles and is used in most simple Monte-Carlo based event generators. Setting the function $f(\mathbf{x,p}) \sim \delta(\mathbf{x} - \mathbf{x(t)})\, \delta(\mathbf{p} - \mathbf{p}(t))$ then just gives the trajectories ($\mathbf{x(t)},\mathbf{p}(t)$), of freely moving particles (in the absence of the collision term).

Together with the frozen approximation this free particle approximation allows to rewrite the collision term from one involving two-body collisions to one that just involves a mean free path in a fixed density. With that ingredient Eq.\ (\ref{Gentransp}) is the one that is being solved by the standard event generators.

\subsection{Factorization in $\nu A$ reactions}
In a neutrino-nucleus reaction the incoming neutrino first reacts with one (or two) nucleons which are bound inside the target nucleus and are Fermi-moving. The reaction products of this very first reaction then traverse the nuclear volume until they leave the nucleus on their way to the detector. This time-development suggests to invoke the so-called factorization of the whole reaction into a first, initial process and a final state interaction process.  The factorization is, however, not perfect. The wave functions of the outgoing nucleons from the first, initial interaction, and thus the cross sections for this initial interaction, are influenced by the potentials present in that final state. The subsequent propagation of particles then takes place in this very same potential.

This is not the case in generators which decouple the first interaction from the final state ones, for example by using different modules for initial interactions and final state propagation, often taken from quite different models. For example, in some versions of GENIE, NEUT and NuWro the spectral functions of nucleons are taken into account in the description of the initial state for the very first interaction. These spectral functions contain implicitly information on potentials and off-shellness.  However, the outgoing nucleons move freely on straight lines, i.e. without experiencing a potential, and only their energies are corrected by some constant binding energy. This is obviously not consistent.  A consistent theory requires to use the full off-shell transport for these 'collision-broadened' nucleons if one is interested not only in semi-inclusive cross sections, but also in quantities such as the final momentum-distribution of the hit nucleon.

While factorization between initial and final state interactions is not exact a careful choice of observables can minimize the coupling between both stages of the collision. This is the central idea behind the proposal by Lu et al \cite{Lu:2015hea,Lu:2015tcr} who have shown that transverse kinematic imbalances of final state leptons and hadrons decouple to some extent from the incoming channel.

\subsection{Preparation of the target ground state}
The nuclear ground state plays an important role for neutrino-nucleus interactions. In most generators it is simply assumed to be described by a free Fermi gas, either global or local. This ansatz neglects all effects of nuclear binding; the nucleus, if left alone, would simply 'evaporate'.

In the nuclear many body approach \cite{Benhar:2006wy} all the effects of a binding potential are contained in the spectral function, even though the potentials themselves cannot be easily determined. The scaling models, on the other hand, do not really need a ground state if the scaling function has been determined from experiment. On the other hand, in the SUSA approach it is explicitly calculated from a relativistic mean field theory \cite{Gonzalez-Jimenez:2014eqa}; its properties then are used in calculating the scaling function. In these calculations the potential is momentum-dependent.

From studies of p-A scattering one knows that the nucleon-nucleus potentials are momentum-dependent such that at small momenta ($< p_F$) they produce binding and at larger kinetic energies of about 300 MeV they disappear \cite{Cooper:1993nx}.
In GiBUU  the ground state is prepared by first calculating for a given realistic density distribution a mean field potential from a density- and momentum-dependent energy-density functional, originally proposed for the description of heavy-ion reactions \cite{Welke:1988zz}.
The potential obtained from it is given by
\begin{equation}  \label{U(p)}
U[\rho,p] = A \frac{\rho}{\rho_0} + B \left(\frac{\rho}{\rho_0}\right)^\tau + 2 \frac{C}{\rho_0} g \int \frac{d^3p'}{(2\pi)^3} \, \frac{f(\vec{r},\vec{p}')}{1 + \left(\frac{\vec{p} - \vec{p}'}{\Lambda}\right)^2}
\end{equation}
which is explicitly momentum dependent. Here $\rho_0$ is the nuclear equilibrium density and $f(\vec{r},\vec{p}')$ is the nuclear phase-space density with $g$ being the spin-isospin degeneracy. If a local Fermi-gas model is used it reads
\begin{equation}
f(\vec{r},\vec{p}') =  \Theta[(|\vec{p}| - p_F(\vec{r})] \quad {\rm with} \quad p_F(\vec{r}) = \left(\frac{6\pi^2}{g}  \rho(\vec{r})\right)^{1/3} ~;
\end{equation}
it is consistent with the spectral phase-space density defined in Eq.\ (\ref{spectrphsp}) for a local Fermi gas momentum distribution, neglecting any in-medium width of the nucleons.  By an iterative procedure the Fermi-energy is kept constant over the nuclear volume; the binding is fixed to -8 MeV for all nuclei. The ground state potential is thus by construction momentum-dependent.
\footnote{The parameters appearing in this potential are given for two different parameter sets in a Table in Ref.\ \cite{Mosel:2018qmv}.} 

This momentum dependence is also present for unbound states and thus affets any processes with outgoing nucleons. In particular, this is also so for QE scattering.

\begin{figure}   
\includegraphics[width=0.9 \textwidth]{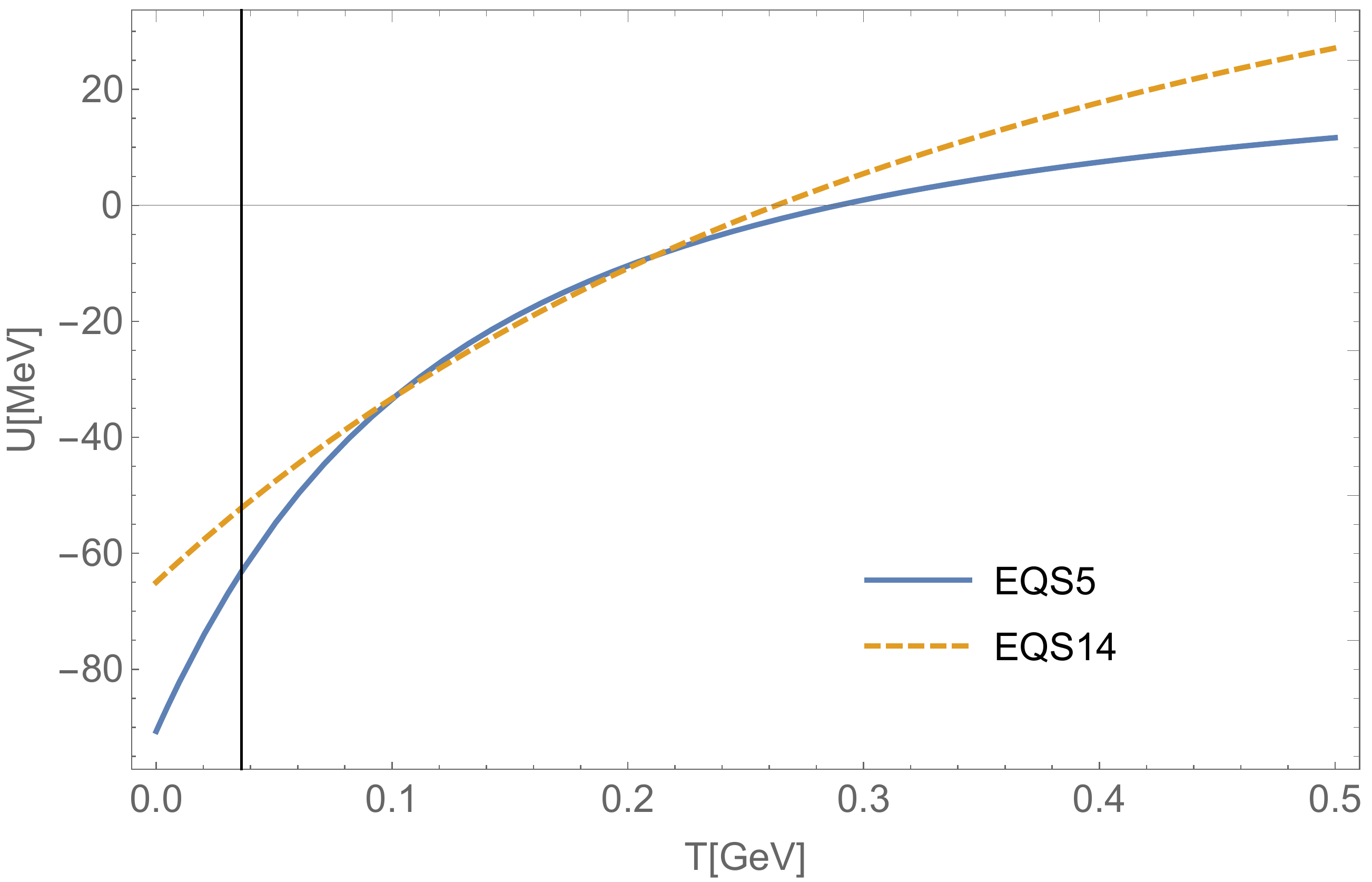}
\caption{The two potentials corresponding to the parametersets EQS5 (solid) and EQS14 (dashed) at density $\rho_0=0.16 fm^{-3}$ as a function of kinetic energy. The solid vertical line indicates the kinetic energy at the Fermi surface. EQS14 (dashed) is obtained from a fit to experimental pA data \cite{Cooper:1993nx} at $r=0$.} \label{fig:Npot}
\end{figure}

Typical potentials are shown in Figure \ref{fig:Npot} in their dependence on momentum $p$.
The potentials EQS5 and EQS14 agree with each other in the kinetic energy range $100 < T < 300$ MeV. For lower kinetic energies, corresponding to lower energy transfers, EQS5 gives a better description of the data (see the results shown in \cite{Gallmeister:2016dnq}). The momentum dependence of the ground state potential and of the potential seen by ejected nucleons is thus consistently obtained from one and the same theory.

In addition to the nuclear potential just discussed for charged particles there is also a Coulomb potential present, which is also implemented in GiBUU, but in none of the other generators. Both potentials together affect mainly low momentum particles. They also cause a deviation of final state particle's trajectories from straight lines which are usually assumed in standard generators.

\subsubsection{RPA correlations}
In calculations that start from an unbound initial state with a local Fermi gas momentum distribution it was found that RPA correlations play a major role in the region of the QE peak \cite{Nieves:2011yp}. Here, at $Q^2 \approx 0.2$ GeV$^2$, uncorrelated calculations were found to describe the data already reasonably well. Adding then RPA effects was found to lower the cross section by about 25\%; after adding then the 2p2h contributions, to be discussed in a later subsection, agreement with the data was again achieved.

\begin{figure}[h]
	\centering
	\includegraphics[width=0.9\linewidth]{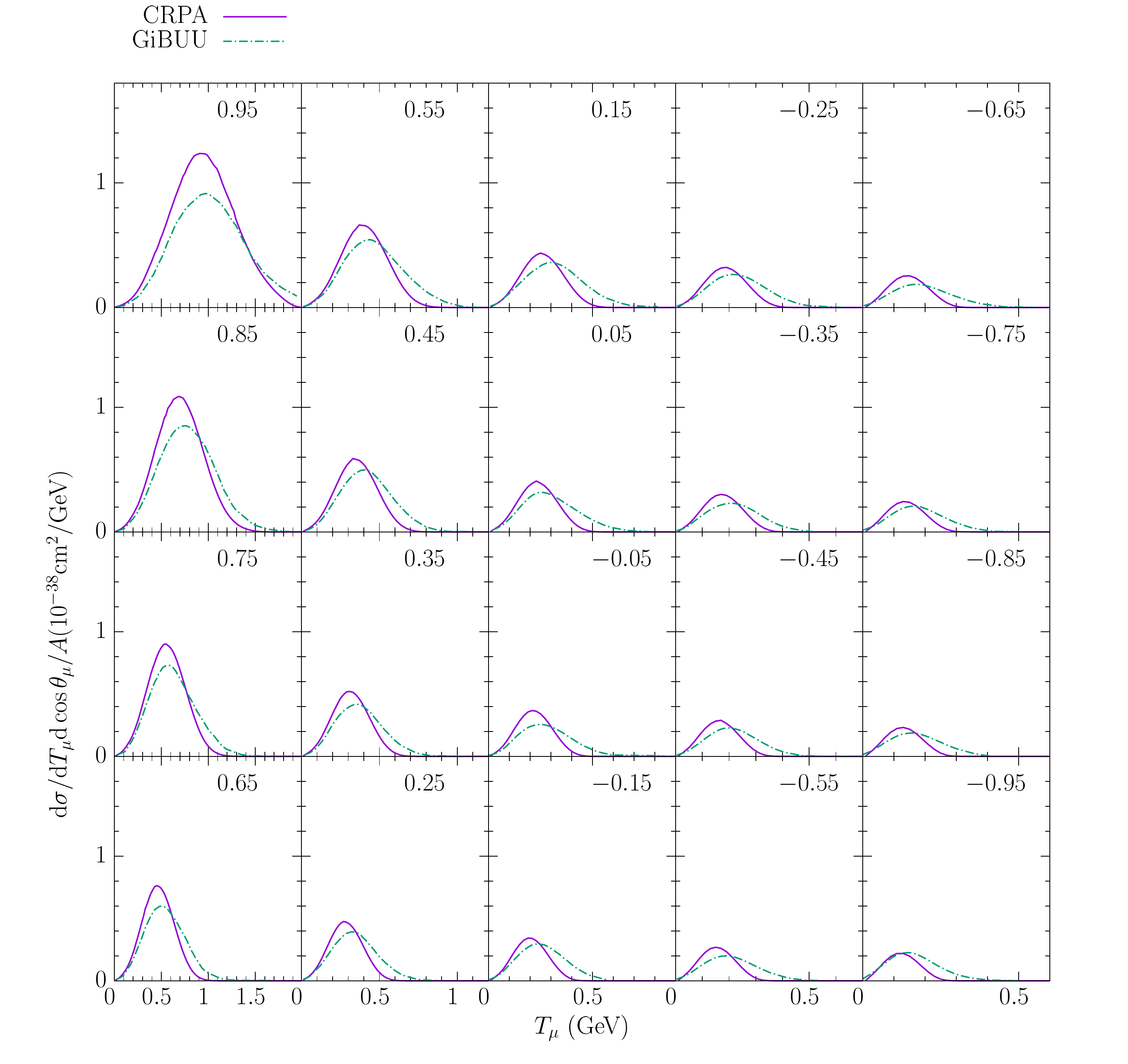}
	\caption{Double-differential QE cross section for outgoing muons in a reaction of neutrinos in the booster neutrino beam with $^{40}Ar$.  The numbers in the upper parts of each subfigure gives the cosine of the muon scattering angle with respect to the neutrino beam. All cross sections are given per nucleon. The solid curve gives the results of a CRPA calculation \cite{VanDessel:2017ery}, the dash-dotted curve those obtained with GiBUU.}
	\label{fig:minervame-Ghent-GiBUU}
\end{figure}

Only recently the Ghent group has shown that this strong lowering caused by the RPA correlations is mostly an artifact due to the use of an unphysical ground state in these calculations \cite{Pandey:2016jju}; this was later also confirmed in an independent calculation \cite{Nieves:2017lij}. In Continuum RPA (CRPA) calculations this group  showed that RPA effects overall are significantly smaller and play a significant role only at small $Q^2$ and small energy transfers ($\approx 10$s of MeV) when a realistic ground state potential is used. This result provides a justification for using mean field potentials in generators without any RPA correlations. Fig.\ \ref{fig:minervame-Ghent-GiBUU} shows a comparison of the double-differential cross section for the QE scattering of the MicroBooNE flux on $^{40}$Ar, calculated on one hand within CRPA \cite{VanDessel:2017ery} and on the other hand -- without any RPA correlations -- within GiBUU. The overall agreement between the models is quite good and illustrates the small influence of RPA correlations
on double differential semi-inclusive cross sections. Since in the calculations of Ref.\ \cite{Nieves:2011yp} the strong RPA effect was nearly canceled by an equally strong, but with opposite sign, effect of 2p2h interactions to reach agreement with experiment one could speculate that also the 2p2h contributions were overestimated in a calculations that starts with a free, non-interacting ground state.

\subsubsection{Spectral functions}
Nuclear many-body theory obtains the hole spectral function $\mathcal{P}_h(\mathbf{p},E)$ (SF) as the imaginary part of the hole propagator: it contains all the information about the energy-momentum distribution of bound nucleons \cite{Benhar:2006wy}. In the semiclassical formalism developed here the same information is contained in $F(x,p)$ at time $t=0$,
i.e.\ at the time of the first contact. The hole spectral function expressed by $F$ is then given by an integral over the nuclear volume
\begin{equation}  \label{Pvolint}
\mathcal{P}_h(\mathbf{p},E) = \int\limits_{\rm nucleus} \!\!\!{\rm d}^3x \, F(\mathbf{x},t=0,\mathbf{p},E) ~.
\end{equation}
The function $F(\mathbf{x},t=0,\mathbf{p},E)$ here can come in principle from any sophisticated nuclear many body theory. It is directly related to the Wigner transform of the one-body density matrix.

In the quasiparticle approximation, relevant for use in generators,the initial $F$ is given by
\begin{equation}  \label{QPapprox1}
F(\mathbf{x},t=0,\mathbf{p},E) = 2 \pi g \, \delta[E - \tilde{E}(\mathbf{x},\mathbf{p})]\, f(\mathbf{x},t=0,\mathbf{p},E) ~
\end{equation}

This quasiparticle approximation -- used together with the global Fermi gas momentum distribution -- has been criticized because it leads to 'spiky', $\delta$-function shaped energy-momentum distributions. Indeed, for free particles without any potential and with a global Fermi gas momentum distribution we have
\begin{eqnarray}
\tilde{E}(x,\mathbf{p}) &=& \sqrt{\mathbf{p}^2 + m^2} \nonumber\\
f(\mathbf{x},t=0,\mathbf{p},\tilde{E}) &=& \Theta\left[p_\mathrm{F} - |\mathbf{p}|\right] \Theta(\tilde{E})
\end{eqnarray}
Then the energy $\tilde{E}$, and therefore also $F$, is no longer dependent on $\mathbf{x}$ so that the spiky behavior is also present after integration over the nuclear volume in Eq.\ (\ref{Pvolint}).

If there is a potential present, however, then the hole spectral function is much better behaved even if the momentum distribution is given by a Fermi-gas approximation. I illustrate this here for a local Fermi gas bound in a scalar potential. The hole spectral function is now given by
\begin{equation}
\mathcal{P}_h(\mathbf{p},E) = 2\pi g \int\limits_{\rm nucleus} \!\!\!{\rm d}^3x \,\Theta\left[p_\mathrm{F}(\mathbf{x}) - |\mathbf{p}|\right] \Theta(E)\, \delta\left(E - m + \sqrt{\mathbf{p}^2 + {m^*}^2(\mathbf{x},\mathbf{p})}\right)~,
\end{equation}
here $p_F$ is the local Fermi-momentum taken to be a function of $p_F(\mathbf{x}) \sim \rho(\mathbf{x})^{1/3}$ (local Fermi-gas) and $E$ is the hole energy taken to be positive. For simplicity it is assumed that in this spectral function all effects of the nucleon potential are contained in the effective mass $m^*$ \cite{Buss:2011mx} which can depend on location and momentum of the nucleon.

The corresponding \emph{momentum distribution} approximates that obtained in state-of-the-art nuclear many-body theory calculations quite well (see Figure 4 in \cite{Alvarez-Ruso:2014bla}); its \emph{energy distribution} no longer contains the $\delta$-function spikes of a free Fermi gas because of the $\mathbf{x}$-dependence of the potential in $m^*$ and the integration over ${\rm d}^3x$. This can be seen in Figs.\ 8 and 9 in Ref.\ \cite{Alberico:1997jg} which show semi-classical spectral functions calculated as discussed above. The SF thus obtained is quite similar to the one obtained from NMBT for the nuclear matter background \cite{Benhar:1994hw} but does not contain the wiggly structure caused by shell effects that are present in the final NMBT SF. These may have an influence on exclusive electron-induced reactions. For neutrino-induced reactions, however, the effect will be minor because of the inherent smearing of all observables over incoming energies due to the flux-distribution of any neutrino long-baseline beam.

Alberico et al.\ \cite{Alberico:1997jg} have given a quite detailed discussion of this semi-classical approach to exclusive electron scattering. The main result of that study was that the potential binding the nucleons has a significant impact on exclusive cross sections which are, in addition, quite sensitive to final state interactions through the mean field. Both of these points are in contrast to the basic assumptions in standard neutrino event generators which neglect mean field potentials both in the initial and the final state of the reaction.

\section{Reaction types}

\subsection{Coherent interactions}
An incoming neutrino can interact with the nucleus in many different ways. It can, first of all, coherently interact with all nucleons. This process happens if the momentum transfer is small and can lead either to elastic neutrino scattering (in a Neutral Current (NC) event) \cite{Akimov:2017ade} or to coherent pion production where the pion carries off the charge of the $W$ in a CC event \cite{AlvarezRuso:2011zz}. In both reaction types the target nucleus remains in its ground state. Such processes can be described by a coherent sum over the individual nucleon amplitudes \cite{Leitner:2009ph} and thus depends crucially on a phase-coherence between all nucleons. This coherence cannot be described by the quantum-kinetic transport equations (or any MC-based generator) which describe the incoherent time-evolution of single particle phase-space densities. Thus, coherent processes fall outside the validity of any semi-classical description and have to be added 'by hand' to any generator.

Theories for the description of coherent processes have been developed for incoming on-shell photons \cite{Peters:1998mb} and experimental results for these are available \cite{Krusche:2002iq}. For neutrino-induced coherent processes the situation is more complicated since experiments usually cannot see de-excitation photons from the target nucleus; one can thus not be sure that the target is left in its ground state. Theoretical investigations of coherent neutrino-induced pion production exploit the fact that the incoming gauge boson 'sees' a coherent superposition of all nucleons and, therefore work with the overall form factor of the target nucleus \cite{AlvarezRuso:2011zz,Hernandez:2009vm}. This is clearly an advantage in particular at higher energies where fully microscopical calculations such as the one in \cite{Leitner:2009ph} are not feasible because of the large number of contributing intermediate states.

\subsection{Semi-inclusive cross sections}
The transport equations describe the full event evolution, from the very first, initial interaction of the incoming neutrino with one (or more) nucleons through the final state interactions up to the asymptotically free final state consisting of a target remnant and outgoing particles.

Lepton semi-inclusive cross sections as a function of the outgoing leptons's energy and angle (or equivalently squared four-momentum transfer $Q^2$ and energy-transfer $\omega$) are then easily obtained by stopping the time-evolution after the first initial interaction. They are defined as the sum of cross sections for all microscopic initial processes. Knowledge about the ultimate fate of the initially struck particles is not required for determining that quantity but their final state wave function enters into the transition amplitude. Such semi-inclusive cross sections obviously are a necessary test for all neutrino generators.

In the theoretical framework outlined above they are obtained by summing over all reaction processes in the first time-step; the further time-development of the reaction is irrelevant for these inclusive cross sections. For example, for the semi-inclusive QE scattering cross section one has
\begin{equation}   \label{inclXsect}
{\rm d} \sigma^{\nu A}_{\rm QE}(E_\nu,Q^2,\omega) =  \int \frac{{\rm d}^3p}{(2\pi)^3} \frac{dE}{2\pi}\,P_h(\mathbf{p},E) f_{\rm corr}\, {\rm d}\sigma^{\rm med}_{\rm QE}(E_\nu,E,Q^2,\omega) \, P_{\rm PB} (\mathbf{p + q})  ~.
\end{equation}
Here ${\rm d}\sigma^{\rm med}$ is a possibly medium-dressed semi-inclusive cross section on a nucleon that depends both on the energy of the incoming neutrino and the energy of the hit nucleon, $f_{\rm corr}$ is a flux correction factor $f_{\rm corr} = (k \cdot p)/(k^0p^0)$ that transforms the flux in the nucleon rest frame into that in the nuclear restrame. The momenta $k$ and $p$ denote the four-momenta of the neutrino and nucleon momentum, respectively and $P_{\rm PB}$ describes the Pauli-blocking and any final state potential or spectral function effects. The cross section ${\rm d} \sigma$ on the lhs of Equation (\ref{inclXsect}) depends on $Q^2$ and $\omega$ and can thus be used to calculate the semi-inclusive cross section.

Final state interactions enter into the inclusive cross sections only through the final states needed to calculate the transition amplitude in the first, initial interaction. The interactions that produced particles experience when they traverse the nucleus are irrelevant.

From these discussions it is clear that all generators, i.e.\ all theories and codes that lead to a full final state event, are also able to describe inclusive and lepton semi-inclusive cross sections. The opposite, however, is not true. Theories that describe the inclusive or semi-inclusive (as function of the outgoing lepton's energy and angle) cross sections very well but inherently integrate over all final momenta of the first interaction do not give any information about the subsequent dynamical evolution of the system. Into this category fall the so-called scaling method and the nuclear-many body theories. Both will be discussed in the following sections.

\subsubsection{Scaling models}
It was observed quite early on that electron scattering data on nuclei show 'scaling' over a kinematical range that roughly covers the low $\omega$ side of the  QE-peak up to its maximum \cite{Day:1990mf}. 'Scaling' here means that the ratios of the nuclear data over the nucleon data in the QE region are described by a function $F(y)$ of the single variable $ y(q,\omega) $ which depends on the momentum transfer $q$ and the energy transfer $\omega$.

Later on, the scaling function was extended to include more detailed information on the particular target nucleus by introducing the Fermi momentum $p_F$ and a shift parameter to fit the strength and the position of the QE peak \cite{Caballero:2010fn}. With these fit parameters an excellent description of electron data in the QE-peak region could be obtained \cite{Antonov:2006md}. This is not too surprising since the width of the QE peak is determined by $p_F$ and its position can be shifted away from the free peak position by a potenial. While all these analyses relied on data models were also developed to calculate the scaling function starting from nuclear theory. This led to the quite successful SUSAv2 model \cite{Gonzalez-Jimenez:2014eqa,Megias:2018ujz} which combines a relativistic mean field description of the target nucleus with a calculation using the Relativistic Plane Wave Impulse Approximation (RPWIA) at higher energies. An open problem here is still the combination of two different models (RMF vs RPWIA) which causes problems for maintaining the relativistic energy and momentum conservation as well as for gauge invariance.

Nevertheless, the SUSA model is an excellent tool to calculate lepton-induced inclusive and semi-inclusive cross sections on nuclei; semi-inclusive here refers to the outgoing lepton. Its main strength lies in the description of semi-inclusive cross sections in the QE region. Inelastic excitations have to be added in by using phenomenological fits to single-nucleon inelastic structure functions. The method does not give any information on the final state of the reaction, except for the outgoing lepton's properties, and thus cannot be used in any generator without invoking further assumptions on the states of the outgoing nucleons.

\subsubsection{Methods from nuclear many-body theory}
Short range correlations in nuclei cause a broadening of the nucleon spectral function which in vacuum is just given by $\delta$ function. Nuclear many body theory (NMBT) has allowed to calculate that function for the nuclear hole states in the nuclear ground state to a high precision \cite{Benhar:1994hw}. By invoking the impulse approximation one then uses this hole spectral function to describe the initial state in a QE scattering event; the final state in this method is usually assumed to be that of a free nucleon (impulse approximation) \cite{Benhar:2006wy}. The calculations thus contain effects in the nuclear ground state that go beyond the mean-field approximation, e.g.\ short-range correlations. The lepton semi-inclusive cross sections are then just given by Eq.\ (\ref{inclXsect}) with the hole spectral function obtained from NMBT by summing the total cross sections for all individual processes. Thus the reliability of this method also depends on the ability of the theory to describe besides QE scattering also pion production (for the T2K energy regime) and also higher-lying excitations and DIS processes for the DUNE flux. In presently available calculations these inelastic excitations are usually taken from fits to inclusive inelastic responses.

So far most of the results are available only for electron-induced reactions \cite{Ankowski:2014yfa}; results for neutrino-induced reactions are mostly missing. Nevertheless, the generators GENIE, NEUT and NuWro have implemented a so-called spectral function option into their description of QE scattering. In this option the very first, initial interaction is described by a cross section for QE scattering obtained by using a spectral function. Such a procedure is dubious since it uses a very different ground state for the QE scattering than for the other processes. Furthermore, in these generators the potential for the outgoing particles is absent and there is, therefore, a discontinuity between the initial state potential, hidden in the spectral function, and the final state potential. The method can thus work only in a limited kinematic range where the outgoing nucleon's kinetic energy is about 250 - 300 MeV so that the momentum-dependent potential nearly vanishes (see Figure \ref{fig:Npot}).

More recently, nuclear many-body theories have had a remarkable success in describing the nuclear ground state and low-lying excitations starting from ab initio interactions \cite{Carlson:2014vla,Lovato:2014eva}. Their main strength lies in calculating lepton semi-inclusive cross sections at relatively low energies since they become the better the closer they stay to the nuclear ground state. There these calculations have the potential to describe not only the inclusive contributions of true one-particle QE events but also admixtures of 2p2h events and short-range correlations (SRC) that could overlap with the former. Indeed, they have recently reproduced experimental results for the semi-inclusive QE response in NC events on light nuclei \cite{Lovato:2014eva}. The method, so far, does not yield any hole spectral functions and can thus not be combined with the impulse approximation, as in the NMBT calculation,s to overcome the limitations of its non-relativistic character. It also does not contain any inelastic excitations nor does it give any information on the final state of the reaction, except for the outgoing lepton's properties, and thus cannot be used in any generator without invoking further assumptions.

\subsubsection{Comparison with experiment}
A quite general difficulty even for describing semi-inclusive cross sections with these methods is that experimental neutrino data always contain a superposition of many different coupled reaction channels. In electron-induced reactions the measurable energy transfer can be used to distinguish at least the bulk parts of QE scattering and resonance excitations, for example. In neutrino-induced reactions, however, the beam energy and thus also the energy transfer is smeared (see discussion in the introduction). Thus, any experimentally observed events in the QE region are always mixtures of different reaction processes with  very similar final, asymptotic states. At the lower beam energy of the T2K experiment, for example, QE and pion production through the $\Delta$ resonance, overlap. This implies that the QE cross section alone cannot be measured in neutrino-induced reactions. This is also true for so-called 0-pion (sometimes also called QE-like) events without any pions in the final state. In this case pions could have been first produced through $\Delta$ excitation and then reabsorbed through FSI. Detailed analyses show that the latter events amount to about 10\% in the T2K energy range ($\approx 700$) MeV \cite{Mosel:2017ssx}. 

Theories, that work well for electrons in describing the semi-inclusive cross sections around the QE peak, thus cannot describe experimental neutrino-induced cross sections, even if these are only for semi-inclusive or 0 pion events.

\subsection{Quasielastic interactions}
The simplest process that can take place in a neutrino-nucleus reaction is that of quasielastic (QE) scattering in which the incoming neutrino interacts with just one nucleon. In theoretical descriptions one usually assumes that the cross section for that process on a Fermi-moving nucleon is the same as that on a free nucleon, except for a necessary Lorentz transformation to the moving nucleon's rest frame ("impulse approximation"). Assuming a Fermi gas momentum distribution for the initial nucleons and free movement for the final state particles allows to give an analytical expression for this cross section \cite{LlewellynSmith:1971zm,Smith:1972xh}. Possible off-shell effects are often treated by a shift of the energy transfer \cite{Benhar:2006wy}

In the early work by Smith and Moniz \cite{Smith:1972xh} there was the assumption hidden in that cross section that the binding energy of the hit nucleon before and after the collision can be neglected; only possible Pauli-blocking of the final state is taken into account. Binding energy effects are then simulated by a change of the energy of the final state nucleon which is assumed to be free \cite{Bodek:2018lmc}; sometimes also the energy transfer is modified \cite{Benhar:2006wy} mainly to correct for the target recoil. Using the so-called impulse approximation then not just inclusive cross sections but also properties of the outgoing nucleons can be calculated.

This impulse approximation makes it possible to use the spectral functions obtained from nuclear many-body theory for a description of the initial state while the final state is still assumed to be free. In terms of the general structure of the transport equations in Section \ref{Generators} this procedure corresponds to a mixture of the full theory as outlined there and the quasiparticle approximation without potentials. The very first interaction is described by the interaction of the incoming neutrino with correlated and bound target nucleons described by broadened spectral functions.  The following transport of the initial final state through the nuclear environment then proceeds as if the nucleons were unbound and $\delta$-function like in their energy-momentum dispersion relation.

In these calculations an initial state potential is inherent in the spectral function even though its precise value is not known. Assuming then a free outgoing nucleon introduces implicitly a momentum-dependence in the potential. From an analysis of $pA$ reactions \cite{Cooper:1993nx} it is known that the nucleon-nucleus potential is momentum dependent such that it is attractive ($\approxeq - 50$ MeV) at low ($< p_F$) momenta whereas it becomes very small at larger momenta ($\approxeq  500$ MeV) (see Figure \ref{fig:Npot}). Such a momentum-dependence of the single-particle potential has been known to affect the position of the QE peak \cite{Rosenfelder:1978qt,OConnell:1988htk}. In \cite{Ankowski:2014yfa} the authors have demonstrated that such a potential has a major influence on the location of the QE peak, in particular at low momentum transfers where the momentum dependence of the potential is strongest. Indeed, at the lowest $Q^2$ the effect of the final state potential, which is introduced from the outside, is dramatic and shifts the QE peak significantly.

\subsection{2p2h processes}
From earlier studies with electrons it was well known that incoming electrons can also interact with two nucleons at the same time in so-called 2p2h processes \cite{Dekker:1991ph,DePace:2003xu}. They tend to fill in the so-called 'dip region' between the QE-peak and the $\Delta$ peak in semi-inclusive cross sections as a function of energy transfer. It was recognized by M. Ericson and her collaborators  \cite{Delorme:1985ps,Marteau:1999jp} that such processes can also play a role in neutrino-induced reactions, in particular if only the outgoing lepton was observed.

This knowledge was rediscovered after the data of the experiment MiniBooNE showed a surplus of so-called quasielastic-like events as a function of reconstructed neutrino energy \cite{AguilarArevalo:2010zc}. The surplus could be explained quite well just by these 2p2h processes \cite{Nieves:2011yp,Martini:2009uj,Martini:2010ex,Martini:2011wp,Nieves:2011pp}. The models used in this work involved various assumptions, such as non-relativistic treatment, unbound local Fermi gas for the ground state and a restriction of the underlying elementary processes to the $\Delta$ resonance region. The latter limits the application of such models to neutrino energies less than about 1 GeV.

In the following subsections some presently used models for the 2p2h contribution in generators are discussed. A common shortcoming of all of them is that they give only semi-inclusive cross sections for the 2p-2h channel. For use in a generator thus an additional assumption has to be made about the momentum-distributions (energy and angle) of the final state particles. Usually a uniform phase-space occupation is imposed which can be formulated very easily in the two-particle center of mass system, followed by a boost to the laboratory system \cite{Lalakulich:2012ac}.

\subsubsection{Microscopic 2p2h contribution}

\paragraph{Lyon-Valencia model}
The calculations first used to explain the MiniBooNE data used various approximations. The calculations reported in \cite{Martini:2009uj,Martini:2010ex,Martini:2011wp} were non-relativistic and involved further approximations such as the neglect of longitudinal contributions to the vector-axial vector interference term. The calculations reported in \cite{Nieves:2011yp}, on the other hand,  were relativistic, but involved approximations in evaluating the momentum-space integrals and in neglecting the direct-exchange interference terms in the matrix elements. Both models start from an unbound ground state assuming a local Fermi-gas momentum distribution. Intrinsic excitations of the nucleon are limited to the $\Delta$ resonance; this limits the applicability of these models to the MicroBooNE, T2K energy regime.

The model does not provide the energy-momentum distributions of outgoing nucleons. For use in a generator it thus has to be supplemented with assumptions about these distributions.

The calculations of the Valencia group \cite{Nieves:2011yp} have found their way into the generators GENIE \cite{Schwehr:2016pvn} and NEUT as options. The model of \cite{Nieves:2011yp} is found to severely underestimate the 2p2h strength in the dip region \cite{Rodrigues:2015hik} which led the MINERvA collaboration to increase the flux-folded strength by multiplying it with  a 2d correction function amounting to an overall factor of 1.53. A comparison of the original Valencia result or the tuned version in GENIE with electron data and with neutrino data from another experiment would be desirable. Contrary to the MINERvA tune this readjustment of the 2p2h strength should take place for the 2p2h structure functions before flux integration.

\paragraph{MEC model}  \label{Megias2p2h} A calculation of the 2p2h contributions that is free of most of the approximations just mentioned was performed in Ref.\ \cite{Simo:2014wka,Megias:2014qva}. These authors evaluated all the relevant diagrams involving 2p2h interactions in the nucleon and $\Delta$ energy regime, using a somewhat simplified $ NN $ interaction. The calculation is fully relativistic and includes all the interference terms. Earlier calculations had assumed that the 2p2h contribution was purely transverse. The new calculations by Megias et al verify that for electrons. For neutrinos they also obtain a longitudinal contribution although the latter is small relative to the transverse one \cite{Megias:2018ujz,Megias:2014qva}. An open problem in these calculations is that only the real part of the $\Delta$ propagator, which appears in all the relevant graphs, is taken into account. The agreement with electron data, usually assumed to be an essential test of an description of neutrino-nucleus data, is quite bad if the full propagator is used.

An advantage of this method is that it predicts the relative ratios of pn pairs vs pp pairs in the outgoing state \cite{RuizSimo:2016ikw}. This is particularly interesting because electron-induced experiments show a clear enhancement of pn vs. pp pair ejection\cite{Duer:2018sxh}. This effect is usually ascribed to the dominance of pn pairs vs pp pairs in nuclear matter. The calculations of RuizSimo et al \cite{Simo:2016imi} show that in a region that is dominated by the meson exchange current (MEC) at least a part of the observed effect is due to the actual interaction process.

The authors have added the microscopic 2p2h cross section calculated as just described to the SUSA description of QE scattering and obtain impressive agreement with semi-inclusive electron and neutrino data \cite{Megias:2016lke,Megias:2016fjk} in the QE and dip region. The MEC model evaluates the two-body current with nucleons up to the $\Delta$ resonance. This is sufficient for the T2K and MicroBooNE energy regime. It does not give any information on the momentum distribution of the outgoing particles and thus cannot be used in a generator without further additional assumptions on the final state of the 2p2h process.

\paragraph{NMBT 2p2h excitations}
In \cite{Rocco:2015cil} the spectral function formalism was extended to include also 2p2h excitations by using standard two-body currents. Results were shown in that paper for electron-induced semi-inclusive cross sections.In a very recent paper \cite{Rocco:2018mwt} results of this method have also been shown for some fixed-energy neutrino reactions. Only semi-inclusive cross sections could be calculated. The results can thus not be used in a generator without further additional assumptions on the final state of the 2p2h process.

\subsubsection{Empirical 2p2h contribution} \label{GiBUU2p2h}
An alternative to the microscopic calculations for 2p2h contributions is to take these directly from an analysis of semi-inclusive
electron scattering data. An analysis by Bosted et al and Christy \cite{Bosted:2012qc,Christy:2015} indeed did extract the structure function for these processes directly from data in a wide kinematical range ($0 < Q^2 < 10$ GeV$^2$, $0.9 < W < 3$ GeV) that goes well beyond the resonance region and implicitly includes any not only MEC, but also SRC and DIS components.

Starting assumption for this extraction was that these 2p2h effects are purely transverse. The parameterized structure function  $W^e_1(Q^2,\omega)$ for electrons thus obtained can then directly be used in calculations of cross sections for electrons. In GiBUU these structure functions have been combined with the other reaction processes and good agreement with semi-inclusive electron data is obtained \cite{Gallmeister:2016dnq}.

Under the assumption that also for neutrinos the dominant 2p2h cross-section contribution ($\sigma^{2p2h}$) is transverse and that lepton masses can be neglected the corresponding cross section can be written in terms of only two neutrino structure functions, $W_1^\nu$ and $W_3^\nu$,
\begin{eqnarray}  \label{LPnu}
\frac{d^2\sigma^{2p2h}}{d\Omega dE'}
&=& \frac{G^2}{2 \pi^2} E'^2 \cos^2 \frac{\theta}{2} \,\left[2W_1^\nu \left(\frac{Q^2}{2\mathbf{q}^2} + \tan^2\frac{\theta}{2}  \right) \right.  \nonumber \\
& & \mbox{}\left. \mp W_3^\nu \frac{E + E'}{M} \tan^2\frac{\theta}{2}\right] ~.
\end{eqnarray}
Here $G$ is the weak coupling constant; $E'$ and $\theta$ are the outgoing lepton energy and angle respectively; $E$ is the incoming neutrino energy;  and $M$ is the nucleon mass.

Walecka et al.\ \cite{Walecka:1975,OConnell:1972edu} have derived a connection between the electron and the neutrino structure functions for 1p processes. In the version used by the Lyon group \cite{Marteau:1999kt} for 2p2h processes it reads
\begin{eqnarray}
\label{eq:W1}
W_1^\nu &=& \left( 1 +  \frac{G_A^2(Q^2)}{G_M^2(Q^2)} \frac{\mathbf{q}^2}{\omega^2} \right)\, W_1^e \,2(\mathcal{T} + 1)
\end{eqnarray}
Here $G_M(Q^2)$ is the magnetic coupling form factor, $G_A(Q^2)$ the axial coupling form factor and $Q^2$ is the squared four momentum transfer $Q^2 = \mathbf{q}^2-\omega^2$.  The structure of $W_1^\nu$ is transparent: to the vector-vector interaction in  $W_1^e$ an axial-axial interaction $~ G_A(Q^2)$ term is added; the axial coupling is related to the vector coupling by an empirical factor $\mathbf{q}^2/\omega^2$ \cite{Marteau:1999kt}. The extra factor 2 is due to the fact that neutrinos are left-handed only. Finally, a factor $(\mathcal{T} + 1)$ appears where $\mathcal{T}$ is the isospin of the target nucleus.

A similar structure shows up in the V-A interference structure function
\begin{equation}   \label{eq:W3}
W_3^\nu = 2 \frac{G_A}{G_M} \frac{\mathbf{q}^2}{\omega^2}\, W_1^e \, 2(\mathcal{T} + 1) ~.
\end{equation}
Exactly this form has also been used in all calculations by the Lyon group \cite{Martini:2009uj}.

The isospin factor in (\ref{eq:W1}) and (\ref{eq:W3}) is derived under the assumption that neutrinos populate the isobaric analogue states of those reached in electron scattering. The Wigner-Eckart theorem then allows to connect the transition matrix elements for electrons ($\sim \tau_3$) with those for neutrinos ($\sim \tau_\pm$). This connection was originally derived by Walecka \cite{Walecka:1975} for single particle processes but it also holds for the 2p2h processes considered here because the relevant transition operators can again be expressed in terms of irreducible tensors of SU(2) \cite{Simo:2016ikv}. As already mentioned this connection depends on the assumption that neutrino processes excite just the isobaric analogues of states reached in electron scattering. It would thus be very interesting to verify the presence of this isospin factor in actual data. So far most of neutrino data were obtained for the $\mathcal{T}=0$ nuclei C and O. It will, therefore, be very interesting to see the effects of this factor for the isospin asymmetric nucleus $^{40}$Ar for which $\mathcal{T}=2$. To experimentally verify this will require a very good knowledge of the incoming neutrino flux and small other uncertainties (see discussion in \cite{Gallmeister:2016dnq,Dolan:2018sbb}).

Equations (\ref{eq:W1}) and (\ref{eq:W3}) relate both the neutrino structure function $W_1^\nu$ and the interference structure function $W_3^\nu$ to just one other function, the structure function $W_1^e(Q^2,\omega)$, determined from electron data. Parameterizing the latter, either by the Bosted et al fit or by some other ansatz, as a function of $Q^2$ and $\omega$ then determines the electron, neutrino and the antineutrino cross sections consistently \cite{Gallmeister:2016dnq}.

This phenomenological model does not make predictions about the magnitude of pp pair vs pn pair ejection since the phenomenological analysis, on which the model is based, did not take any final state information into account. Instead, in GiBUU which uses this phenomenological description, the isospin composition of pairs is entirely determined by statistical ratios.

Because of the wide kinematical range of the data that were used to extract $W_1^e$ the model is applicable to experiments with high incoming energy (MINERvA, NOvA, DUNE). Since it is based on an empirical analysis of semi-inclusive 2p2h processes the model does not give any information on the final state. In GiBUU it is, therefore, supplemented with the assumption that the energy- and momentum-distributions of the two outgoing nucleons are determined by phase-space.

\subsection{Pion production}
In neutrino-nucleus reactions pion-production plays a major role. In particular at the higher energies of the MINERvA or DUNE experiment pion-production, through resonances or DIS, makes up for about 1/2 - 2/3 of the total interaction cross section and even at T2K it amounts to up to 1/3 of the total. Therefore it must be under quantitative control in the generators used to extract cross sections and neutrino mixing properties from such experiments. Very recently, two comparisons of the generators NEUT (with charged pion - nucleus data) \cite{PinzonGuerra:2018rju} and GENIE (with MINERvA neutrino-induced pion production data) \cite{Stowell:2019zsh} have shown that major discrepancies of these generators with the data exist that cannot be tuned away.

Neutrino-induced pion production on nuclei can - similar to the QE scattering description - be treated by describing the pion production on single nucleons that are bound and Fermi-moving. For the lower energy transfers pion production proceeds predominantly through the $\Delta$ resonance. The binding energy correction is then usually handled by assigning a omplex density-dependent selfenergy to the $\Delta$ which takes care of such effects as Pauli-blocking and collisional broadening.
The pion production cross section in the resonance region is then obtained as a coherent sum over resonance and background amplitudes \cite{Leitner:2006ww,Leitner:2006sp,Hernandez:2010bx}. The models of Leitner et al \cite{Leitner:2006ww,Leitner:2006sp} and Hernandez \cite{Hernandez:2010bx} are essentially identical and differ only in their treatment of the background amplitude.

\subsubsection{Resonance amplitudes}
The resonance amplitudes are determined by the nucleon-resonance transition currents; the corresponding interaction vertices involve form factors. The number of these form factors is tied to the spin of the resonances. Thus, spin-1/2 resonances are connected with two form factors and spin 3/2 ones, such as the $\Delta$ resonance, with four such form factors. This is true for electromagnetic interactions where only vector couplings appear. For neutrinos, in addition axial couplings are present which require the same number of formfactors again. Thus, for the case of a spin-isospin 3/2-3/2 resonance as for the $\Delta$, there are four vector form factors and four axial ones.

 In standard neutrino generators GENIE and NEUT  the so-called Rein-Segal model \cite{Rein:1980wg} for the form factors for resonance excitation is still used although that model is known to fail in its description of electron scattering data \cite{Graczyk:2007bc,Leitner:2008fg}. It is not used in any electron-induced studies of nucleon resonances.  

A better way is to obtain the vector form factors $C_i^V(Q^2)$ (i=3,...,6), which are directly related to the electromagnetic transition form factors \cite{Leitner:2009zz}, from the measured helicity amplitudes (see \cite{Leitner:2009zz}). The helicity amplitudes are determined in, e.g., the MAID analysis \cite{Drechsel:2007if}; the connection between these helicity amplitudes and the vector form factors is given in \cite{Leitner:2009zz}. Current conservation here imposes a constraint on one of the vector form factors to vanish ($C_6^V=0$). This is the approach followed in GiBUU.

For the axial form factors the situation is less well determined since the quality of these data is not sufficient to determine all four axial form factors $C_i^A(Q^2)$. Already in Ref.\ \cite{{Albright:1965zz}} it was noticed that $C_5^A$ gives the dominant contribution. $C_6^A$ can be related to $C_5^A$ by PCAC \cite{Lalakulich:2005cs} leaving also in the axial sector only three form factors. In addition, $C_3^A$ is set to zero based on an old analysis by Adler \cite{Adler:1968tw}, whereas $C_4^A$ is linked to $C_5^A$. Based on these relations all theoretical analyses have so far used the  axial form factor $C^A_5(Q^2)$ (see Eq.\ (18) in \cite{Leitner:2006ww}) with various parameterizations. The latter usually go beyond that of a simple dipole \cite{Leitner:2006ww,Leitner:2006sp,Lalakulich:2005cs,Hernandez:2010bx} and have been obtained by fitting the neutrino pion-production data on an elementary target available from two experiments performed in the middle 80's at Argonne National Laboratory (ANL) \cite{Radecky:1981fn} and at Brookhaven National Laboratory \cite{Kitagaki:1986ct}.

Both experiments also had extracted various invariant mass distributions from their data. The analysis of these invariant mass data together with the experimental $d\sigma/dQ^2$ distributions then led the authors of \cite{Lalakulich:2010ss} to conclude that probably the BNL data \cite{Kitagaki:1986ct} were too high. This has been confirmed by a reanalysis of the old data by Wilkinson et al \cite{Wilkinson:2014yfa} who used the QE data obtained in the same experiment for a new flux calibration. After that flux recalibration the BNL data agree with the ANL data within experimental uncertainties. There remains an uncertainty, however, that is connected with possible final state effects in the extraction of pion production cross sections on the nucleon from data obtained with a deuterium target \cite{Wu:2014rga}. In a very recent calculation that covers a wider kinematical regime Nakamura et al have tried to correct the old bubble chamber data on pion production for Fermi motion and final state effects \cite{Nakamura:2018ntd}. They find in particular at lower neutrino energies corrections of the order of 10 - 20 \%. New, more precise measurements on elementary targets are needed to verify this result.

\subsubsection{Background amplitudes}
A complication in determining the form factors for neutrino-induced excitations of nucleon resonances from comparison with data is due to background contributions that contribute to the observed cross section. For \emph{electro}-production of pions $t$-channel processes provide a background to the resonance contribution. The total cross section is then given by the coherent sum of $s$-channel and $t$-channel amplitudes. Analogously, also for the case of \emph{neutrino}-interactions there is a  background contribution due to Born-type diagrams where the incoming $W^+$ (for neutrino-induced CC pion production) interacts with the nucleon line $W N \rightarrow N' \pi$. The Lagrangian for this latter interaction can be obtained from effective field theory, for low energies up to about the $\Delta$ region \cite{Lalakulich:2010ss,Hernandez:2007qq,Alvarez-Ruso:2015eva}. The description for the $\Delta$ region obtained with this model is quite good, but for the higher-lying resonances one has to resort to modeling both the resonance and the background contributions. Since for higher lying resonances there is even less experimental information available all models simply use dipole parameterizations for the resonance transition form factors with the strength obtained from PCAC \cite{Lalakulich:2010ss}.

Much more ambitious, but also significantly more involved, is the dynamical coupled-channel model of photo-, electro- and weak pion production developed in Ref.\ \cite{Nakamura:2015rta} that has been applied to all resonances with invariant masses up to 2.1 GeV. In this model background and resonance contributions emerge from the same Lagrangian and thus the relative phase between resonance and background amplitudes is fixed. Furthermore, not only pion, but also other meson production channels, for example, for the important 2 pion production as well as for kaons and etas, can be predicted. These elementary data on the nucleon are an essential input into calculations for production on nuclei. 

Representing a coherent sum of amplitudes in a semiclassical, fundamentally incoherent generator poses a problem that requires some practical approximation. In GiBUU, for example, the resonance part alone is handled by exciting nucleon resonances that are then being propagated as new particles until they decay again. The sum of the background part and the interference part is lumped together into a background term; it is assumed that pions that are due to this 'background' are produced immediately. GENIE uses instead the Rein-Sehgal model for the resonance excitation and adds a tuned fit to average total cross sections as background. NuWro uses a similar procedure.

\subsubsection{Pion production and absorption}
Pion production and pion absorption are closely linked through basic quantum mechanical constraints. This can be clearly seen for an energy regime where only the nucleon resonances are essential and DIS does not yet contribute significantly, i.e.\ in the energy regime of T2K and MicroBooNE. Here the resonance ($\Delta$) contribution to pion production proceeds via
\begin{equation}   \label{Wpiprod}
W^+ + N \rightarrow \Delta \rightarrow \pi + N' ~,
\end{equation}
whereas pion absorption proceeds through the same resonance
\begin{equation}    \label{piabs}
\pi + N \rightarrow \Delta \qquad \Rightarrow \qquad \Delta + N \rightarrow N' + N'' ~.
\end{equation}
In both processes the very same $\pi N \Delta$ vertex appears, once in the $\Delta$ production and once in its decay.

The generators GENIE, NEUT and NuWro all violate this time-reversal invariance condition by using quite different models for pion production and absorption. For example, in these standard generators  the production is described by resonance excitations within the Rein-Segal model, whereas pion reabsorption is handled by a very different model, mostly the pion-absorption cascade of the Valencia group \cite{Salcedo:1987md}. This is particularly dangerous if then -- as usual in the generators -- tuning parameters are introduced that allow to tune pion production and pion absorption independently from each other. This obviously introduces artificial degrees of freedom. The amount of 'stuck-pion events', i.e.\ events in which a pion was first produced and then reabsorbed in the same target nucleus, are most sensitive to any imbalance between pion production and pion absorption. The generator NuWRo uses a model with a formation-time parameter for the emitted pion even in the  $\Delta$ resonance region \cite{Golan:2012wx}. This gives additional tuning degrees of freedom for pion absorption which, however, have no physical basis: In the resonance region the time-development of pion production is governed by the resonance width; there is no room for additional parameters.

Finally, it is worthwhile to point out that descriptions of pion production in generators can use many available data on photo- and electro-production of pions on nuclei for a check of the method \cite{Krusche:2004uw,Krusche:2004zc,Clasie:2007aa,ElFassi:2012nr}. Unlike any other generator GiBUU, which respects the stringent connection between pion production and pion absorption in the resonance region, has been applied to many of them \cite{Hombach:1994gb,Effenberger:1996im,Effenberger:1999jc,Lehr:1999zr,LehrDiss:2003,Muhlich:2004zj}. For neutrino-induced reactions the GiBUU results for pion production are generally in agreement with experimental data \cite{Gallmeister:2016dnq,Mosel:2017ssx,Lalakulich:2010ss,Mosel:2015tja,Mosel:2017nzk,Mosel:2017zwq}.

\subsection{Deep inelastic scattering}
Neutrino physicists have traditionally defined all reactions connected with the emission of more than one pion as Deep Inelastic Scattering (DIS). This is an oversimplification since over the last 30 years the study of nucleon excitations has shown that up to invariant masses of about 2 GeV there are many resonances that also decay into two and even more pions; the 2 pion threshold opens at a mass of about 1.5 GeV. Only above about 2 GeV the individual nucleon resonances start to overlap and the DIS regime starts. Furthermore, nucleon resonance excitations and DIS have very different $Q^2$ dependencies such that DIS is connected with events with $Q^2 > 1$ GeV$^2$.

The semi-inclusive cross sections for DIS can be expressed in terms of structure functions \cite{Leader:1996}. In the pQCD regime, i.e.\ $Q^2 > 1$ GeV$^2$, these structure functions can be written down in terms of parton distribution functions. In regions, where pQCD does not yet provide the correct description, parameterizations of these structure functions have been obtained by fits to semi-inclusive cross section data. The high-energy event generator PYTHIA \cite{Sjostrand:2006za} is then often used to actually provide also mass- and energy-distributions of the final state. When using this framework for neutrino-nucleus reactions one is faced with the complication that the target nucleon is bound, i.e.\ off shell. Various schemes have been investigated to deal with this problem; it was found that these effects play only a small, but visible role at intermediate energies \cite{Lalakulich:2012gm}.

Also for DIS reactions electro-production data provide an excellent testing ground for generators. In particular the HERMES data, taken with a lepton beam at 28 GeV and 12 GeV incoming lepton energy on nuclear targets up to Xe \cite{Airapetian:2007vu,Airapetian:2011jp}, but also the JLab data taken by Brooks et al \cite{Accardi:2009qv} are particularly relevant for such tests. These data have been analyzed with an early version of GiBUU \cite{Gallmeister:2007an}. This latter study also has shown that the often used prescription to forbid any interactions within a so-called 'formation-time' is unrealistic and not in agreement with the data from HERMES and the EMC experiment.

\section{Final state interactions}
In the preceding sections I have already mentioned the importance of final state interactions (FSI), e.g.\ in connection with the final state potential in QE processes or in connection with pion reabsorption. These final state interactions are due to hadron-hadron interactions. They are thus independent of the electroweak nature of the initial interaction. In this section I now summarize some of the methods used in generators to describe FSI. This description can necessarily only be rather superficial since there often exist no detailed write-ups of the physics used in these generators and their algorithms. A notable exception is GiBUU for which both the physics and many details of the numerical implementation have been extensively documented \cite{Buss:2011mx}.

Final state interactions can be split into different categories:
 
\begin{itemize}
\item The final state wavefunction for the very first, initial interaction is affected by a potential for the outgoing hadrons. This FSI obviously affects the initial transition rate.

\item 
This same potential acts not only in the initial interaction but also during all of the following cascade. Particles then move on possibly complicated trajectories that can have an influence on observed final state angular distributions.

\item
Another kind of FSI is that which an initially produced particle experiences when it collides with other nucleons inside the target on its way out
of the nucleus to the detector. In such collisions both elastic and inelastic scattering, as well as particle production, can take place, possibly connected also with charge transfer. The cross sections for these processes can be affected by the nuclear medium with the changes depending on the local density and the momentum of the interacting particles \cite{Li:1993rwa}.

An often not discussed problem is connected with the final state of the target remnant. The nucleons in the target nucleus carry energy and momentum which have to enter into the overall energy-conservation of the primary interaction, both in the initial and the final state. This also holds for a final state collision between the primarily ejected nucleon and a target nucleon, but in the usually used frozen configuration this is often not taken into account.

\end{itemize}

One problem that one faces in this description is that the neutrino 'illuminates' the whole target nucleus and, therefore, any kicked out (nucleons) or produced (e.g.\ pions) particles can start their way through the nuclear target at any density. This is very different from a reaction in which an incoming particle hits a nucleus from its outside. For example, $\pi A$ absorption data are sensitive to the overall absorption, they do not give, however, any information on the mean free path inside the nucleus as long as their mean free path is smaller than the nuclear diameter. A significantly better check is provided by electro- or photo-production of pions on nuclei ($\gamma A \to \pi A^*$) since the incoming photons illuminate the whole nuclear volume, just as neutrino do. Unfortunately, no such comparisons of standard generator results with photo-production data are available, the notable exception being again GiBUU \cite{Krusche:2004zc,Krusche:2004uw}.

In the following discussion I will briefly go through some of the models used in generators\footnote{For GENIE I rely on the manual, dated March 13, 2018, to be found at https://genie-docdb.pp.rl.ac.uk/cgi-bin/ShowDocument?docid=2}.
\begin{enumerate}
\item {\bf Effective Models}
   \begin{itemize}
      \item GENIE hA
	   In this model the cascade is reduced to a single step in which e.g.\ pions are impinging on an iron nucleus. Their absorption is calculated and then scaled to other nuclei.
      \item GENIE hA 2014 uses the same oversimplification, only the $A$-scaling has been improved
   \end{itemize}
   This model, which still is the default in GENIE goes back to the INTRANUKE program developed about 25 years ago; it is "simple and empirical, data-driven" in the words of the GENIE manual, but is really quite outdated and misses the relevant physics: a particle + A reaction does not describe the particle being produced inside the nucleus and then cascading through it.
\item {\bf Cascade models}
    \begin{itemize}
    \item  GENIE hN is a " full intranuclear cascade" model, according to the manual of v. 3.0, but no details are given. Obviously pion production and absorption are treated independent from each other. Hadrons move freely, without potentials.
      \item NuWRo The treatment of FSI in NuWRo is similar to that in GENIE hN and thus the same criticisms applay.
    \item  FLUKA \cite{FLUKA} is a model that has been widely used for all sorts of nuclear and hadron interaction studies. It also has a neutrino option \cite{Battistoni:2009zzb}, but so far has not been used to analyze or reproduce neutrino data from ongoing experiments.
    \item NEUT Nucleon beam scattering data are used to tune the nucleon FSI. The final state interactions of different particles, such as pions, are handled by introducing tunable multiplicative factors that are different for different phyics processes. For pions the FSI were originally handled by simple attenuation factors. More modern versions of the code use the Valencia cascade for the FSI of pions; this violates detailed balance as discussed above.
    \end{itemize}
\item {\bf Transport Models}
    The GiBUU transport model starts the outgoing particles inside the nuclear volume at their point of creating and then propagates them out of the nucleus. It respects the detailed balance constraints on pion production and absorption in the resonance region. The relevant cross sections for both processes are calculated within one and the same theoretical model (see App. B1 in \cite{Buss:2011mx}). The pion final state interactions are then handled by a full, relativistically correct cascade. Calculations for neutrino-nucleus reactions so far have used the frozen configuration approximation. GiBUU does allow also to treat the target nucleons dynamically; this could become important when at high incoming energies the target nucleus is significantly disrupted.

\end{enumerate}

\section{Tuning of generators}
Finally, a comment is in order on tuning the generators to data. This is a widespread practice among experimentalists using neutrino generators such as GENIE and NEUT. Cross sections and potentials are crucial inputs to all generators and they all carry with them some experimental uncertainties; varying them within their experimental error bars is well justified. Such fits of physics parameters to data could actually help to decrease the experimental uncertainties on elementary in-medium cross sections.

On the other hand, often also unphysical parameters are tuned. An example is the use of different momentum distributions (global vs. local Fermi gas) in describing different elementary processes or the 'brute force' change of physics input. An example for the latter is provided by the tuning of 2p2h cross sections in the MINERvA experiment\cite{Rodrigues:2015hik}. The GENIE version used for that analysis originally started with a description of 2p2h processes taken from the Valencia model \cite{Nieves:2011pp}. After comparing with data it was found that this 2p2h contribution provided by that model is too weak and its integrated strength had to to be increased by about 53\% \cite{Ruterbories:2018gub}. This increase was not achieved by changing coupling constant or similar physics parameters, but instead by fitting a 2d function with free parameters directly to the data.

A very recent analysis of the MINERvA pion production data has shown that even allowing tuning of GENIE ingredients does not give a satisfactory fit to the data \cite{Stowell:2019zsh}. Similarly, an analysis of pion-nucleus data using the generator NEUT has shown significant discrepancies between data and generator result \cite{PinzonGuerra:2018rju}. One may wonder, however, about the ultimate conclusion from these studies which probably just showed that incorrect physics models could not even be fitted to data. GiBUU, on the other hand, which contains a consistent pion production model in the resonance region, has sucessfully reproduced many pion data without any special tune \cite{Mosel:2017ssx,Mosel:2017zwq}.

\section{Generator results}
In this section I use some results of the generator GiBUU to illustrate some properties of neutrino-nucleus interactions in different energy regimes and for different targets. I will start with semi-inclusive cross sections and then also show some more exclusive observables for the experiments T2K, MINERvA, MicroBOONE and DUNE.

All results shown in this section were obtained with the 2019 version of GiBUU \cite{gibuu} running in the quasiparticle approximation described in Sect.\ \ref{sect:QPapprox}. No special tune was used; all calculations for different incoming flux distributions and for different targets were obtained with one and the same set of input parameters 'out of the box'.

\subsection{Comparisons with electron data}
Even though the community in general agrees that checks of generators against electron data are necessary there exist very few published comparisons of generator results with such data. For NEUT there are no published comparisons available, for GENIE only very recently some comparisons have become available \cite{Ankowski:2019, Ashkenazi:2018} which indicate serious discrepancies with data in all of the inelastic excitation region. For NuWro only some preliminary results exist \cite{Zmuda:2015twa}. For GiBUU there is a long list of comparisons both with electron and photon data available \cite{Mosel:2018qmv,Gallmeister:2016dnq,Gallmeister:2007an,Effenberger:1996rc,Lehr:1999zr,Lehr:2001an,Lehr:2003ka,Lehr:2003km,Buss:2006vh,Buss:2006yk,Buss:2007ar,Buss:2008sj,Gallmeister:2010wn,Hombach:1994gb,Muhlich:2003tj,Muhlich:2002tu,Kaskulov:2008ej,Kaskulov:2011ab} which cover both inclusive and semi-inclusive particle production data obtained in electro-nuclear and photo-nuclear reactions.

\subsection{Semi-inclusive cross sections}

\subsubsection{T2K}
\begin{figure}[ht]
	\centering
	\includegraphics[width=0.8\linewidth]{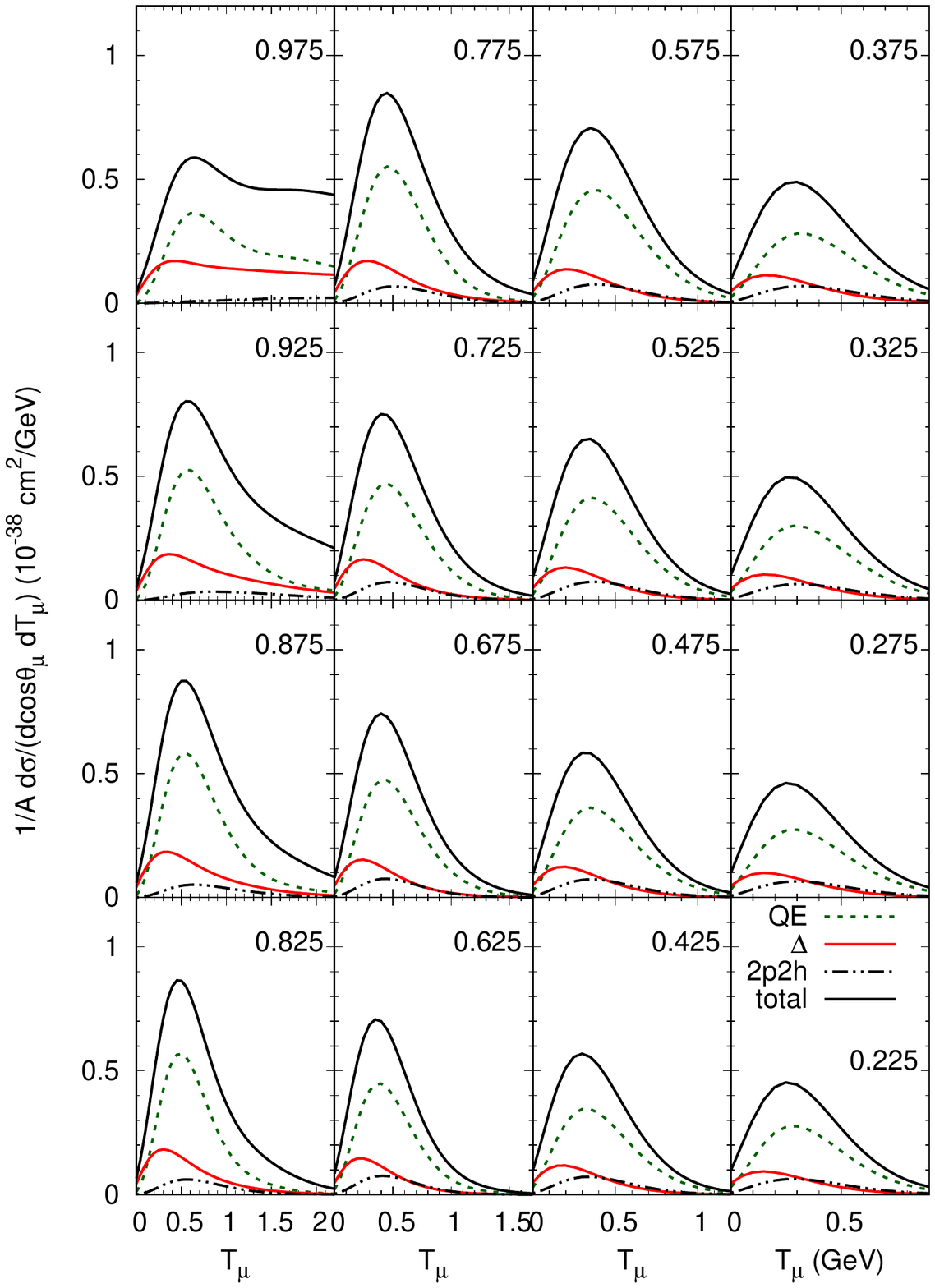}
	\caption{Double-differential cross section per nucleon for outgoing muons for the T2K neutrino beam hitting a $^{12}C$ target at the near detector. The different curves depict the contributions of CCQE scattering, $\Delta$ excitation and 2p2h processes to the cross section as a function of the outgoing muon kinetic energy, as indicated in the figure. The numbers in the upper parts of each subfigure gives the cosine of the muon scattering angle with respect to the neutrino beam. All cross sections are given per nucleon.}
	\label{fig:T2K-incl}
\end{figure}

The mean beam energy at T2K is about 650 MeV so that one expects that here QE scattering and pion production through the $\Delta$ resonance are dominant. This is indeed borne out in the semi-inclusive cross section shown in Fig.\ \ref{fig:T2K-incl}.
While all angular bins look very similar, with a QE scattering peak at about $T_\mu = 0.5$ GeV, the most forward bin ($\cos\,\theta=0.975$) shows a different behavior, with a long, flat shoulder out to higher $T_\mu$. This behavior shows up already in the individual QE and $\Delta$ contributions shown separately in the figure. It is due to the higher energy tail in the incoming neutrino flux. In addition, also DIS starts to contribute at about $T_\mu = 1$ GeV (not explicitly shown in the figure). Note that the 2p2h contribution is essentially absent in the most forward bin. This reflects the transverse character of this reaction type in GiBUU.

\subsubsection{MicroBooNE}
\begin{figure}[ht]
	\centering
	\includegraphics[width=0.8\linewidth]{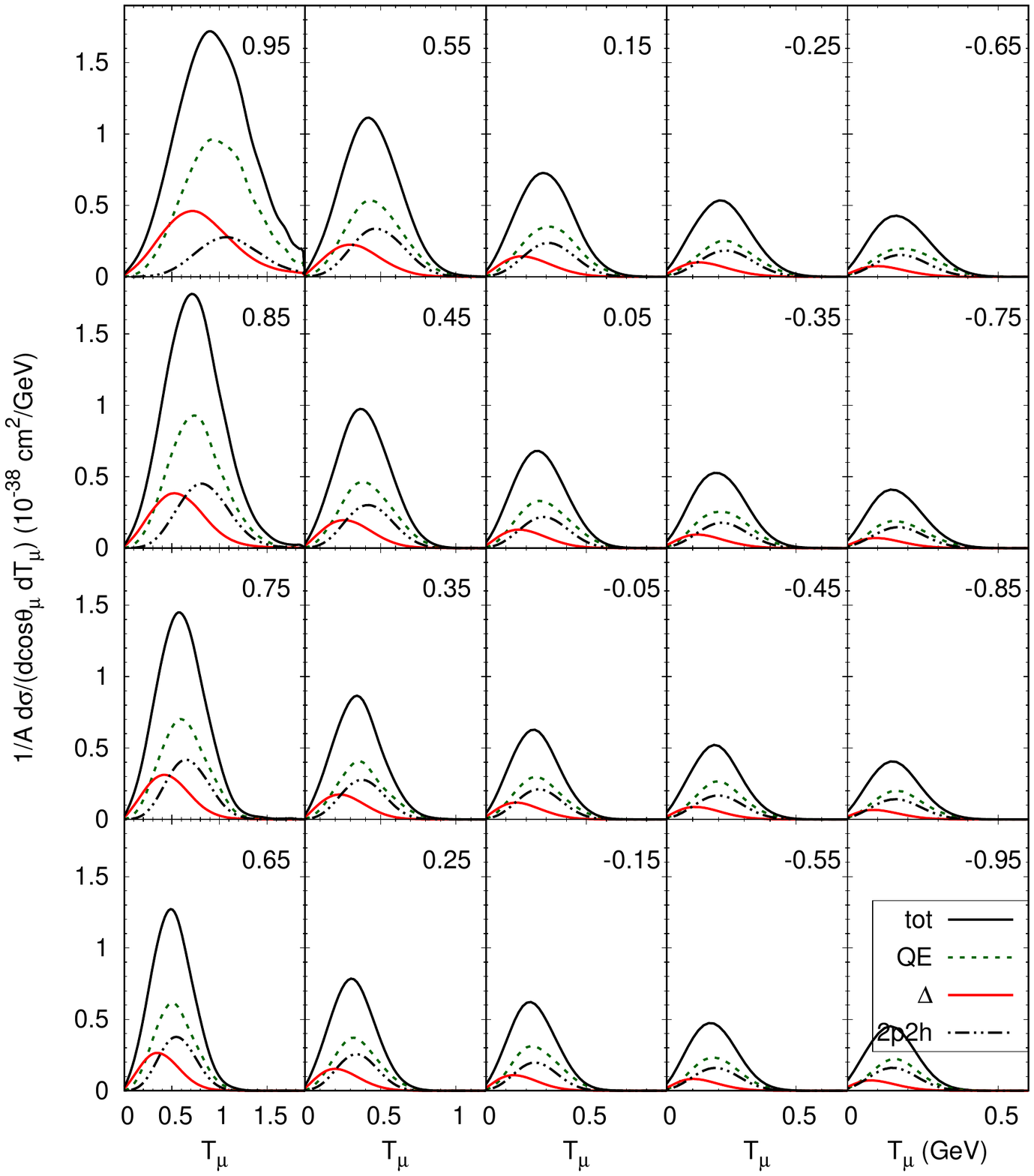}
	\caption{Double-differential cross section per nucleon for outgoing muons for the MicroBooNE experiment with a $^{40}Ar$ target. The different curves depict the various contributions to the cross section as a function of the outgoing muon kinetic energy, as indicated in the figure. The numbers in the upper parts of each subfigure gives the cosine of the muon scattering angle with respect to the neutrino beam. All cross sections are given per nucleon.}
	\label{fig:microb-incl-multiplot}
\end{figure}

MicroBooNE is an experiment that runs in the Booster Neutrino Beam with an $^{40}$Ar target. It has $4\pi$ coverage; therefore I show in Fig.\ \ref{fig:microb-incl-multiplot} the double differential cross section over the full angular range.
The cross section is now presented in wider bins than the one shown before for the T2K experiment. As a consequence, the first bin is centered less forward. The flat behavior of the total cross section in the forward bin for T2K thus does not show up here in the plot of the double differential distribution for MicroBooNE. Otherwise, the results are quite similar, with the exception of the 2p2h contribution that is now larger than for T2K, reflecting the non-zero isospin of the target nucleus $^{40}$Ar ($\mathcal{T}=2$). It will be most interesting to see if this increase in the 2p2h strength is indeed borne out by the data that are presently being taken \cite{Adams:2019iqc}. A verification would give direct information on the
states that are excited in a neutrino-nucleus 2p2h reaction since the isospin factor emerges under the assumption that neutrino-induced reactions populate the isobaric analogue states of electron scattering experiments.

\subsubsection{MINERvA LE and ME}
\begin{figure}[ht]
	\centering
	\includegraphics[width=0.8\linewidth]{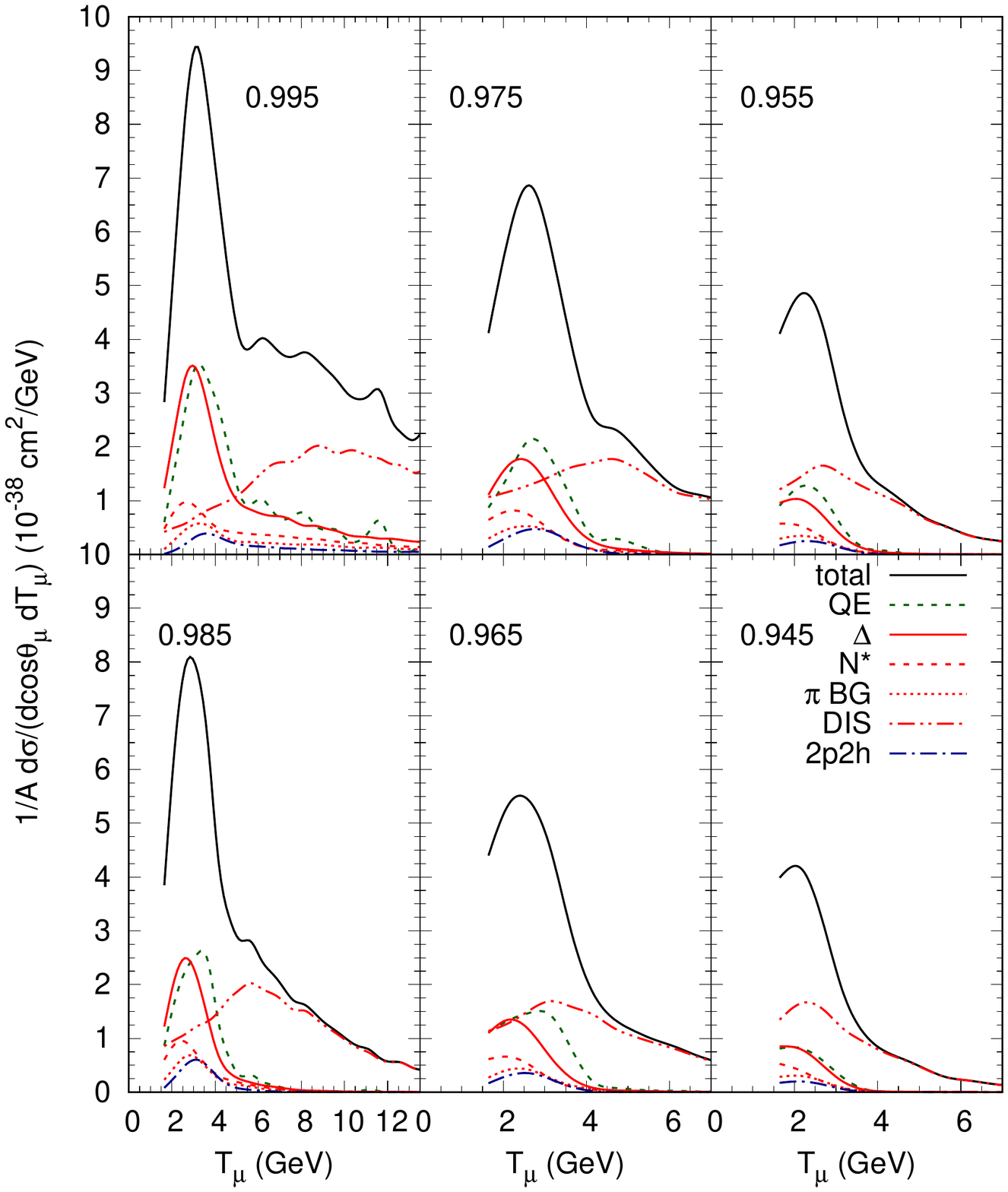}
	\caption{Double-differential cross section per nucleon for outgoing muons for the MINERvA lower energy neutrino beam hitting a $^{12}C$ target. The different curves depict the various contributions to the cross section as a function of the outgoing muon kinetic energy, as indicated in the figure. The numbers in the upper parts of each subfigure gives the cosine of the muon scattering angle with respect to the neutrino beam. All cross sections are given per nucleon.}
	\label{fig:minervale-incl-multiplot}
\end{figure}

The MINERvA experiment has run at a higher energy, but with targets similar to those used in the T2K ND detector. The mean beam energy in its lower energy (LE) configuration is about 3.5 GeV so that one expects a larger contribution of resonance excitations and even DIS in this case. This can indeed be seen in the lepton semi-inclusive double-differential cross sections as a function of muon kinetic energy in various angular bins in Fig.\ \ref{fig:minervale-incl-multiplot}. The experiment has acceptance cuts. Muons with energies below about 1.5 GeV and with angles larger than 20 degrees are not detected. These cuts are also used in the figure.

One sees that the cross sections are strongly forward peaked. In the most forward bin ($\cos\,\theta=0.995$) QE and $\Delta$ excitation nearly completely overlap. Both together make up more than 3/4 of the total cross section in that bin. The 2p2h contribution is comparatively negligible (about 5\% of the total); at the peak of the total cross section the 2p2h contribution is the smallest of all. DIS accounts for less than 10 \% in the peak region of that bin, but it has a long high-energy tail\footnote{In GiBUU all events connected with nucleon excitations above an invariant mass of 2 GeV are identified as 'DIS'}. For muon kinetic energies above about 4 GeV the DIS contribution becomes dominant. In the next angular bin ($cos\,\theta=0.985$) DIS accounts for nearly all of the cross section for $T_\mu > 4$ GeV. This defines an optimal kinematical region for studies of DIS in neutrino-induced reactions.

\begin{figure}[ht]
	\centering
	\includegraphics[width=0.7\linewidth]{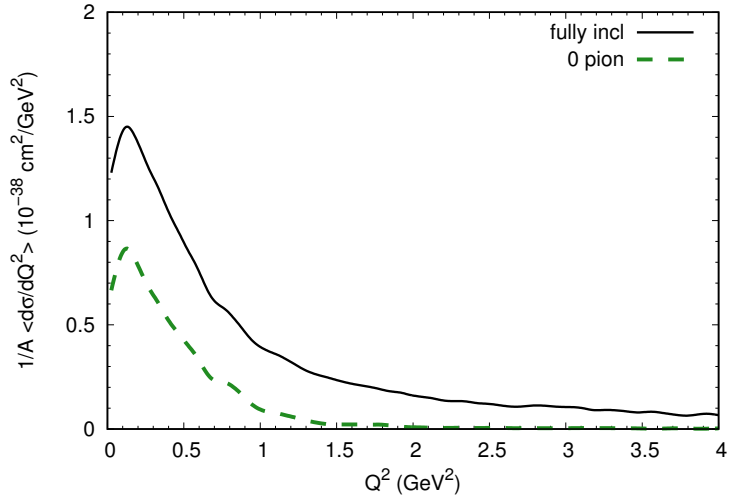}
	\caption{$Q^2$ distribution per nucleon for inclusive and for 0-pion events in the MINERvA LE beam hitting a $^{12}C$ beam.}
	\label{fig:minervadsigmadq2LE}
\end{figure}

Figure \ref{fig:minervadsigmadq2LE} gives the $Q^2$ distribution, where $Q^2$ here is calculated from
$Q^2 = 4 E_\nu E_\mu sin^2(\theta_\mu/2)$ ~.
Since $Q^2$ cannot directly be measured, but must be reconstructed, this is a less direct observable than the double differential cross sections. Nevertheless it is interesting to see that the inclusive $Q^2$ distributions reach out to fairly large $Q^2$ while the distribution for 0-pion events dies out at $Q^2 \approx 1.5 \:{\rm GeV}^2$. The latter reflects the fact that in 0-pion events resonance excitations and DIS, which both dominantly decay into a nucleon and pions, are suppressed. DIS events are connected with momentum transfers $Q^2 > 1$ GeV${^2}$ where they contribute significantly to the inclusive cross section.

\begin{figure}[ht]
	\centering
	\includegraphics[width=0.8\linewidth]{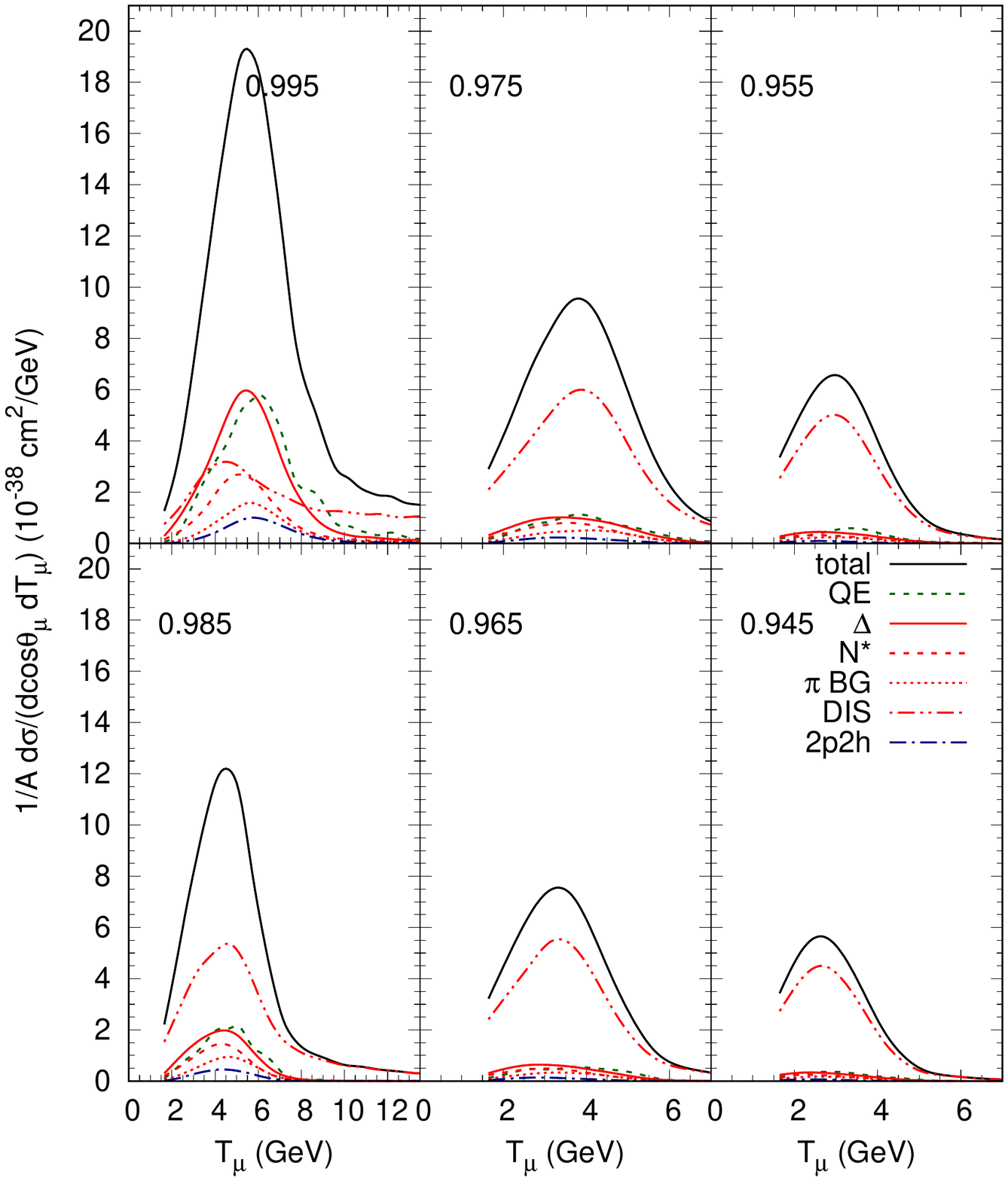}
	\caption{Same as Fig.\ \ref{fig:minervale-incl-multiplot} for the MINERvA medium energy neutrino beam.}
	\label{fig:minervame-incl-multiplot}
\end{figure}

\begin{figure}[ht]
	\centering
	\includegraphics[width=0.7\linewidth]{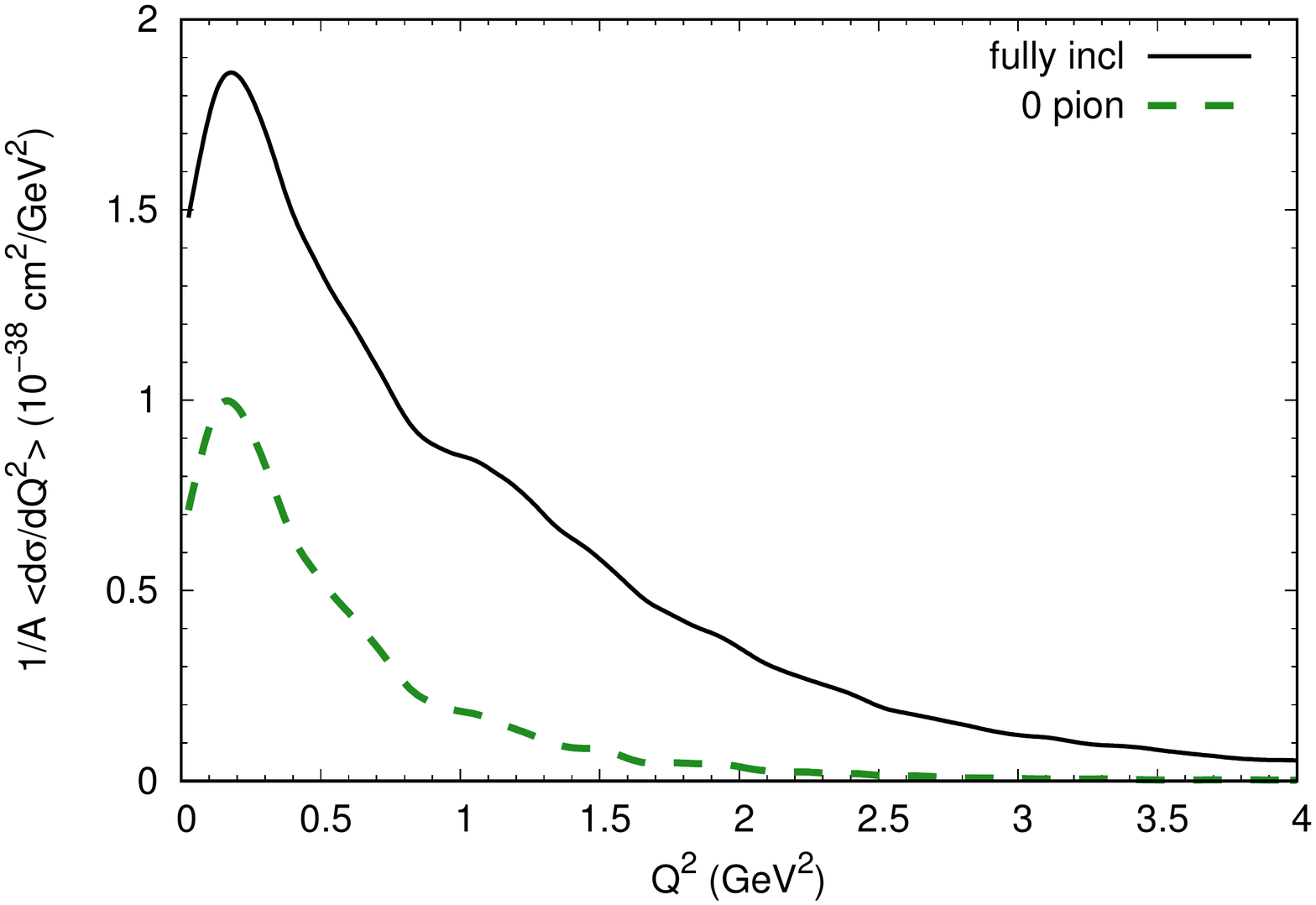}
	\caption{$Q^2$ distribution per nucleon for inclusive and for 0-pion events in the MINERvA ME beam hitting a $^{12}C$ target.}
	\label{fig:minervadsigmadq2ME}
\end{figure}

Presently, the MINERvA experiment is also analyzing data from a so-called medium energy (ME) run where the flux peaks at about 5.75 GeV. If one looks at the same distributions as before now for the medium energy energy beam in Fig.\ \ref{fig:minervame-incl-multiplot} one sees similar shapes for the overall cross sections. Again $\Delta$ excitation and QE scattering give about equal contributions to the cross section at the most forward bin. In that bin ($\cos\,\theta = 0.995)$ the DIS contribution is larger already under the peak, but its shoulder is not so visible as in Fig.\ \ref{fig:minervale-incl-multiplot} because the other components are broader. It becomes dominant in the next angular bin ($\cos\,\theta = 0.985$) and from $\cos\,\theta = 0.975$ on it accounts for nearly all the cross section. Choosing only events with the outgoing muon angle $> 10$ degrees thus enriches the DIS events significantly.
The $Q^2$ distribution (Fig.\ \ref{fig:minervadsigmadq2ME}) is similar to the one at the LE flux. Again restricting the events to those with zero outgoing pions cuts the cross section by nearly a factor 2 at small $Q^2$ and brings the cross section down to nearly zero at $Q^2 \approx 2.5$ GeV.

\subsubsection{ArgoNeut}
\begin{figure}[ht]
	\centering
	\includegraphics[width=0.8\linewidth]{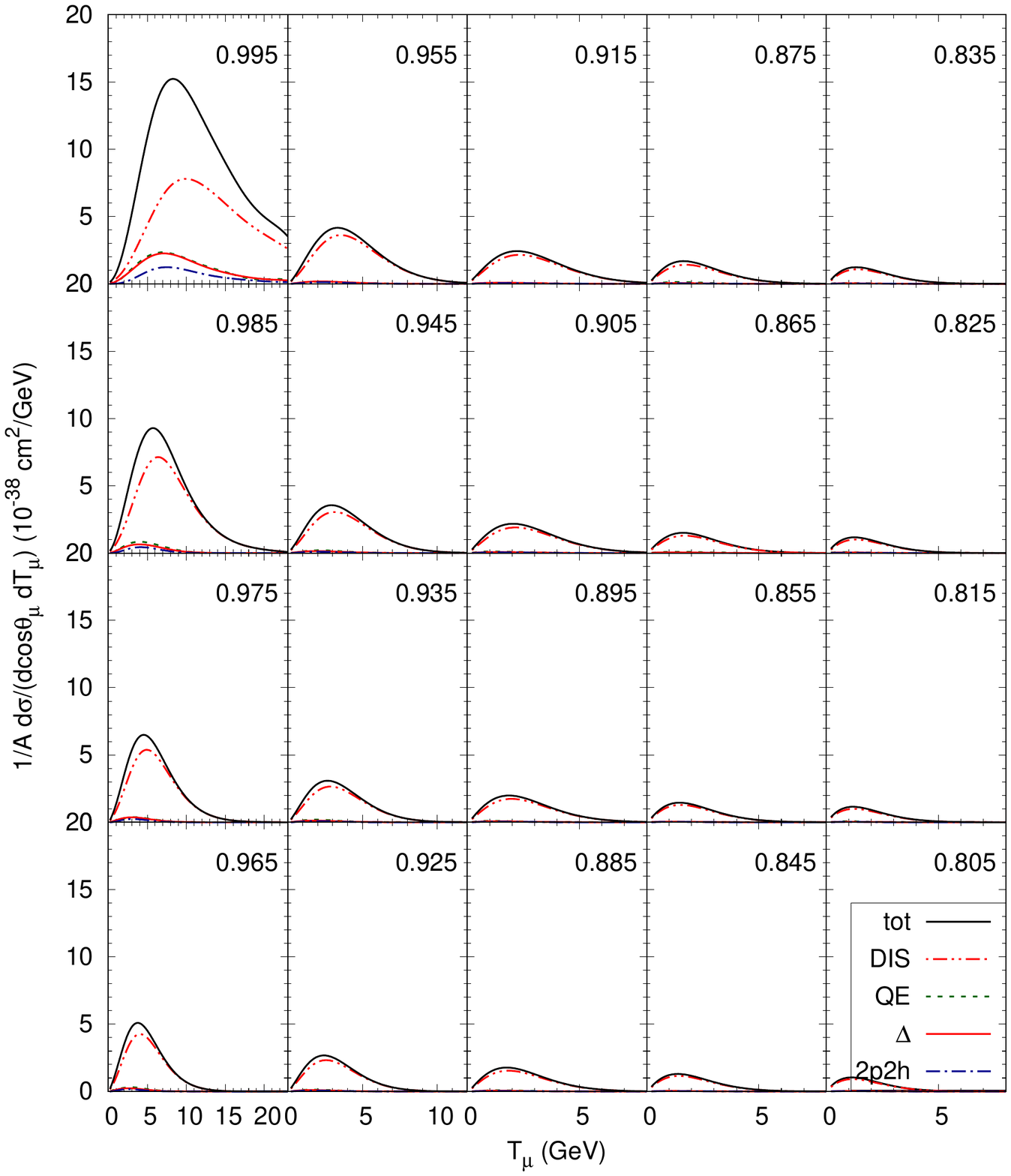}
	\caption{Lepton semi-inclusive double differential cross section in the forward region for the ArgoNEUT experiment. The numbers in each bin give the central $\cos\,\theta$ value of the angular bin. Some contributions from different reaction mechanisms are indicated in the figure. }
	\label{fig:Argodd}
\end{figure}

The data from ArgoNeut are particularly interesting because this experiment was the first to use an Ar target in a higher energy beam. Fig.\ \ref{fig:Argodd} shows the double-differential cross section for a neutrino beam; this flux peaks at about 6 GeV with a long, hard tail out to larger energies.
It is noticeable that now even in the most forward bin ($\cos\,\theta = 0.995$) the largest contribution comes from DIS, which contributes about a factor of three more than QE and $\Delta$ processes. This is due to the fact that DIS cross sections depend approximately linearly on the incoming neutrino energy and the ArgoNeut flux has significant strength also at higher energies. These Ar data could in principle serve as a testing ground for the isospin dependence of the 2p2h component. However, this component is so small even for $\mathcal{T}=2$, compared to the dominant DIS contribution, so that there is hardly any sensitivity to its strength.

\begin{figure}[ht]
	\centering
	\includegraphics[width=0.8\linewidth]{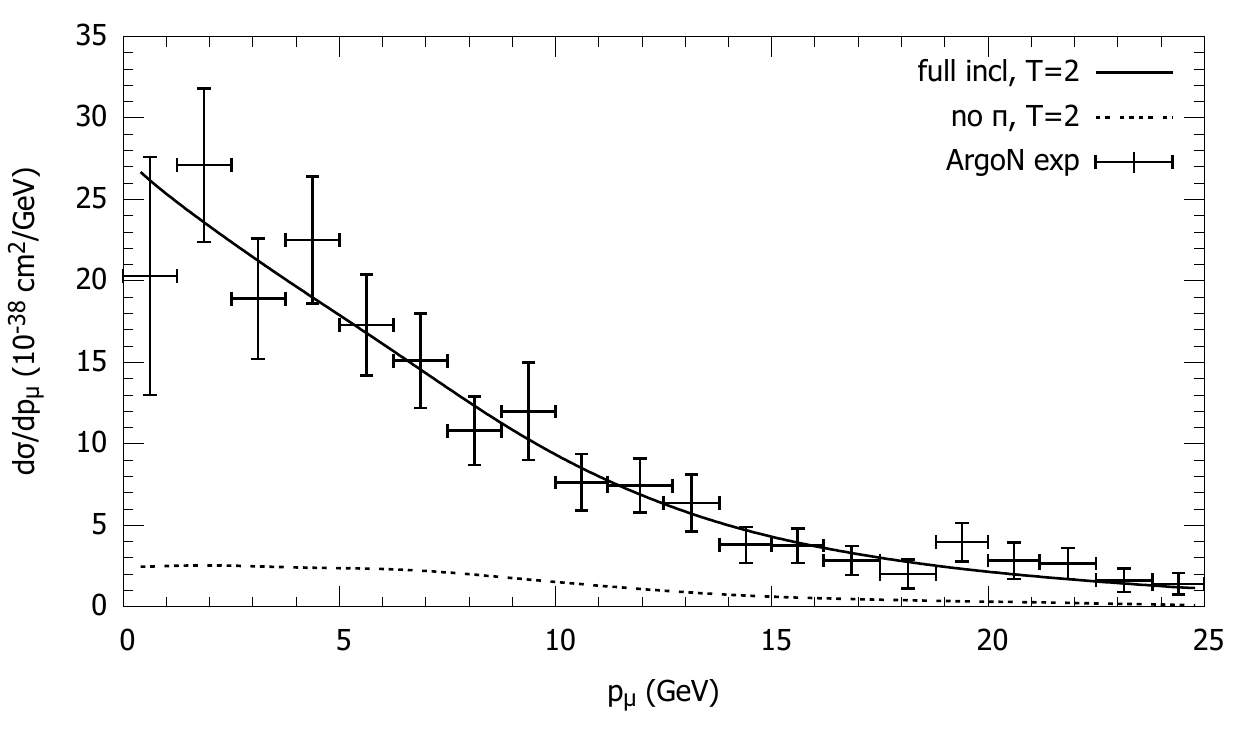}
	\caption{Semi-inclusive muon momentum spectrum for the ArgoNEUT experiment. The dashed curve gives cross section for $0 \pi$ events. Data are from \cite{Acciarri:2014eit}.}
	\label{fig:Argopmu}
\end{figure}

Fig.\ \ref{fig:Argopmu} gives the muon momentum spectrum for the same experiment. Shown are results for a calculation with $\mathcal{T}=2$, but a calculation with $\mathcal{T}=0$ gives nearly the same curve because the overall contribution of 2p2h processes is small. Also shown is the spectrum for 0-pion events. This cross section is quite flat and significantly smaller than the  inclusive one. The latter feature is due to the dominance of DIS in the fully inclusive cross section and to the fact that DIS events nearly always lead to pions in the final state.

\subsubsection{DUNE}
Finally we discuss here the semi-inclusive cross section for the DUNE near detector (ND). DUNE will work with a neutrino flux similar to the MINERvA LE flux, but with $^{40}$Ar as a target. On the one hand, we thus expect a very similar behavior as for MINERvA, in particular a strong forward peaking of the cross section. On the other hand, the isospin of $^{40}$Ar is $\mathcal{T}=2$; this leads to an overall enhancement of the 2p2h cross contribution by a factor 3. It will be interesting to look for any observable consequences of that isospin change.

\begin{figure}[ht]
	\centering
	\includegraphics[width=0.8\linewidth]{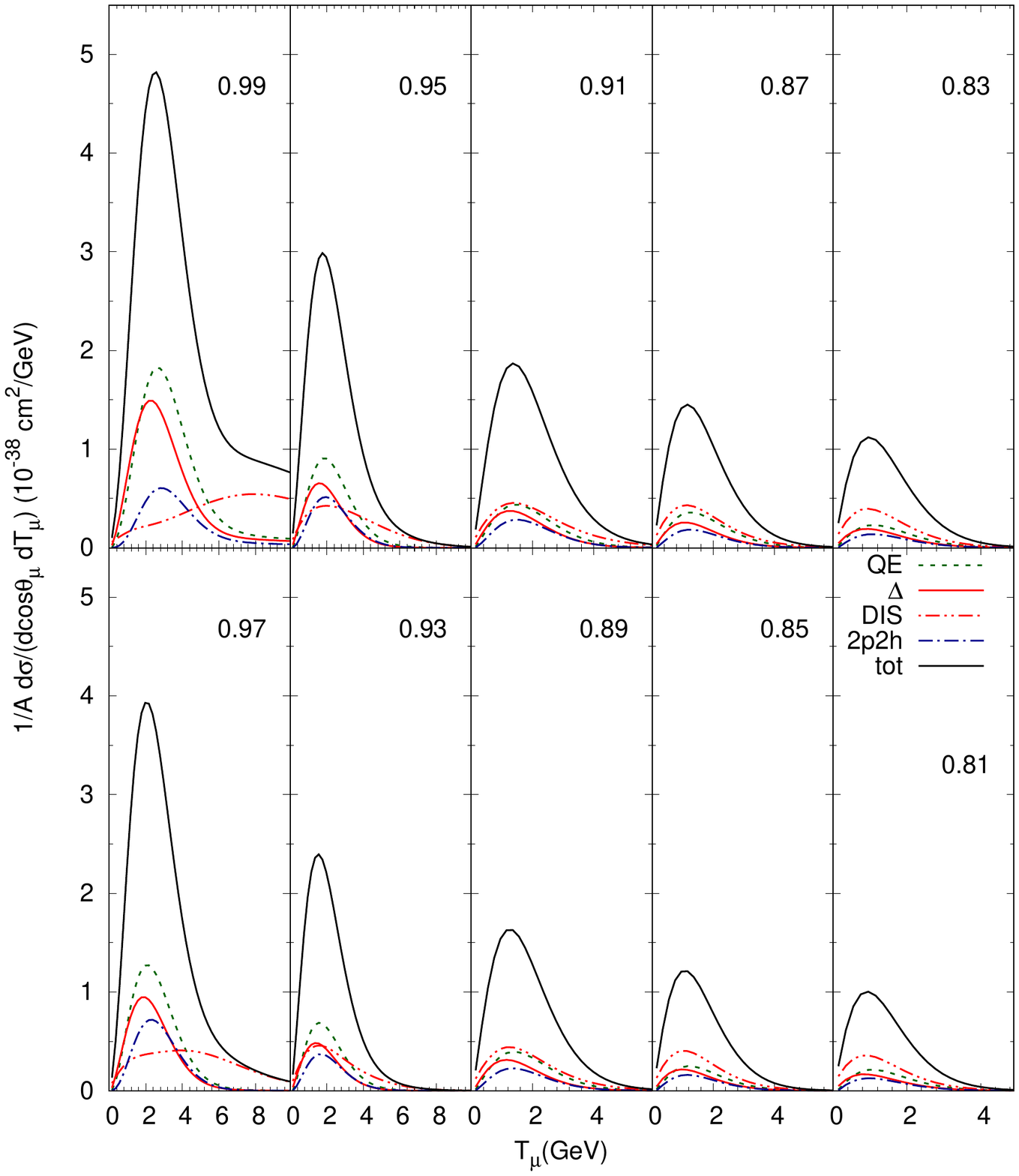}
	\caption{Same as Fig.\ \ref{fig:minervale-incl-multiplot} for the DUNE near detector in the 2017 flux.}
	\label{fig:DUNEincl}
\end{figure}

All calculations for the DUNE ND were performed with the flux from \cite{DUNEFlux}. That the double differential cross section is indeed quite similar to the one obtained for the MINERvA LE run can indeed be seen in Fig.\ 15 of Ref.\ \cite{Gallmeister:2016dnq}. We show, therefore, now only the most forward angles in Fig.\ \ref{fig:DUNEincl}.
Most noticeable is the significantly larger (than in the MINERvA LE run) contribution of 2p2h excitations that comes about because the target nucleus Ar has $\mathcal{T}=2$ whereas in the MINERvA experiment the target is C with $\mathcal{T}=0$. This difference leads to an enhancement of the 2p2h cross section by a factor 3 relative to the one on C. DIS is not as large as it is in the ArgoNeut experiment because the incoming flux does not have the sizeable high energy tail that the ArgoNEUT flux had.

\subsection{Semi-inclusive cross sections for hadrons}
From the preceding discussions it is clear that already the lepton semi-inclusive cross sections are made up of various quite different reaction mechanisms, the most important ones being QE scattering and pion production (through resonances and DIS). This is already a challenge even for theories that describe inclusive cross sections, such as GFMC and SUSA inspired models, since they usually cannot handle the elastic and the inelastic processes without further approximations.

In this subsection I will now illustrate some features of semi-inclusive cross sections for processes with outgoing hadrons in the DUNE Near Detector (ND). These can, for example, be spectra of certain given particles while an integration over all other degrees of freedom is performed. This is the actual strength of generators such as GiBUU which yield information about the complete final state.

\subsubsection{Transparency}
\begin{figure}[ht]
	\centering
	\includegraphics[width=0.8\linewidth,]{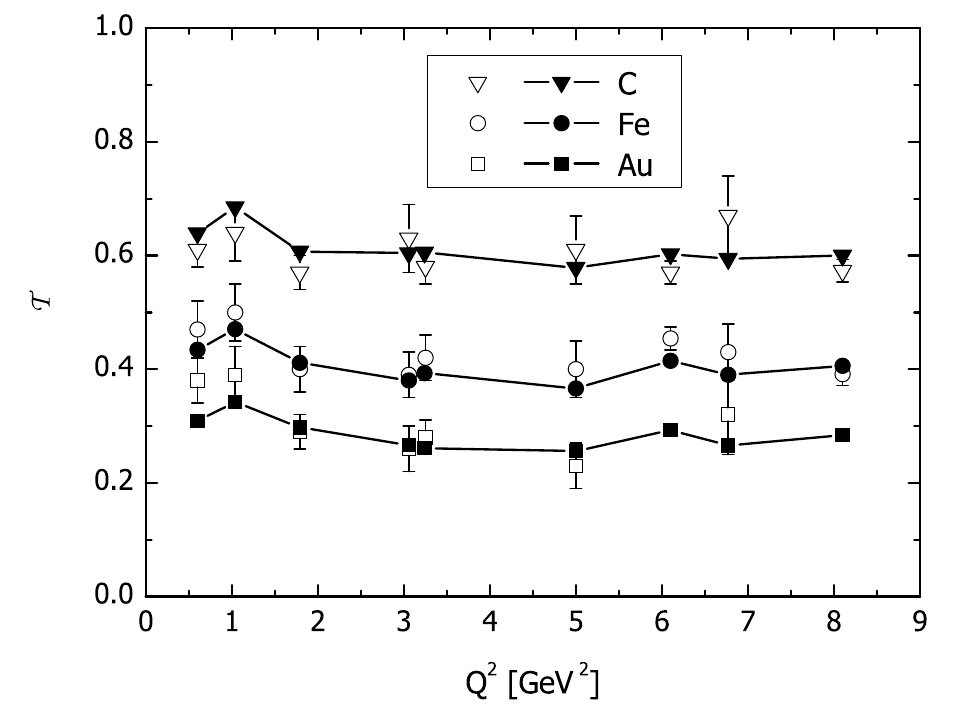}
	\caption{Transparency ratio for C, Fe and 
	Au compared to data (open symbols) from JLAB \cite{Abbott:1997bc,Garrow:2001di} and SLAC \cite{ONeill:1994znv}. The full symbols represent the GiBUU results. Taken from \cite{Lehr:2001an,LehrDiss:2003}.}
	\label{fig:transp}
\end{figure}

Before coming to the DUNE cross sections for outgoing nucleons it is interesting to look at a comparison with the transparency of nuclei for outgoing nucleons that was measured in electron-induced reactions for different targets and over a wide range of momentum-transfer $Q^2$. Experimentally, the transparency was defined as a ratio of the  number of nucleons in a given angular interval, chosen to be approximately symmetric around the momentum of the virtual photon, to the same number expected if there were no final state interactions. In the experimental work the latter number was obtained from Glauber or Distorted Wave Impulse Approximation calculations. The theoretical transparencies were, on the other hand, consistently obtained by running GiBUU with and without final state interactions. The comparison of GiBUU results, obtained already in 2001,  with the data is shown in Fig.\ \ref{fig:transp}.

The agreement is obviously quite good. Both the $Q^2$-dependence and the $A$-dependence are described very well in a calculation that used no 'unusual' effects, such as color transparency. The structures in these curves are mostly explained by experimental constraints on the intervals over which the experimental transparencies were determined \cite{LehrDiss:2003}, the peak at $Q^2 \approx 1$ GeV$^2$ reflects a minimum in the nucleon-nucleon interaction cross section at a proton momentum of about 0.7 GeV. A more detailed discussion of there results can be found in \cite{LehrDiss:2003}.

\subsubsection{Particle spectra}
\begin{figure}[ht]
	\centering
	\includegraphics[width=0.8\linewidth]{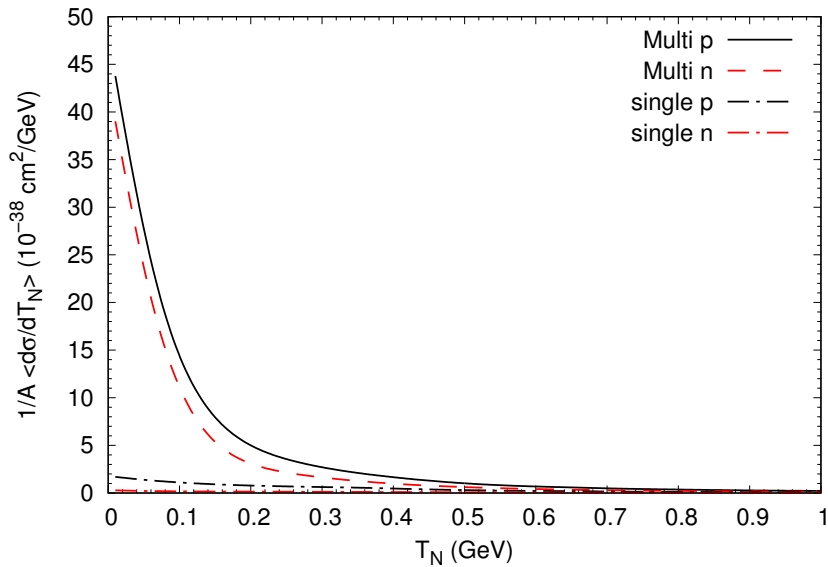}
	\caption{Kinetic energy spectra for protons and neutrons in the DUNE ND. Shown are results both for single and for multinucleon events. The single events contain one and only one nucleon of the given isospin. The Multi events contain at least one nucleon of the given isospin and any number of nucleons with the same or a different isospin.}
	\label{fig:DUNEdsigmadtn}
\end{figure}

\paragraph{Nucleons}
Fig.\ \ref{fig:DUNEdsigmadtn} shows the kinetic energy spectra of protons and neutrons expected in the neutrino beam for the DUNE experiment. Plotted are both the spectra for events with one and only one outgoing nucleon and those for multi-p or multi-n events\footnote{The cross section shown in Fig.\ \ref{fig:DUNEdsigmadtn} for Multi events is that for events with one of the nucleons with the given isospin and others with any isospin present. The kinetic energy is that of any one nucleon with the given isospin in such an event}. While the former are small and quite flat as a function of nucleon kinetic energy the multi-nucleon events exhibit a steep rise in their spectra below about 0.3 GeV \footnote{Kinetic energy below about 0.02 - 0.04 GeV cannot be trusted in a semiclassical theory because of general quantum-mechanical effects becoming essential \cite{Buss:2006vh}.}. This rise is due to final state interactions: While at the end of the first, initial neutrino-nucleus interaction there may be only one nucleon outgoing, its collisions with other nucleons causes an 'avalanche' of nucleons. Energy conservation then requires that all these secondary particles carry lower energies.
This steep rise also implies that caution has to be taken when a calorimetric energy reconstruction is performed. Typically, the detectors have a lower detection threshold of about 50 MeV \cite{Acciarri:2015uup}. This means that a large part of the nucleon ejection cross section is not visible. This is even more so in detectors that do not see outgoing neutrons in the final state.

\begin{figure}[ht]
	\centering
	\includegraphics[width=0.8\linewidth]{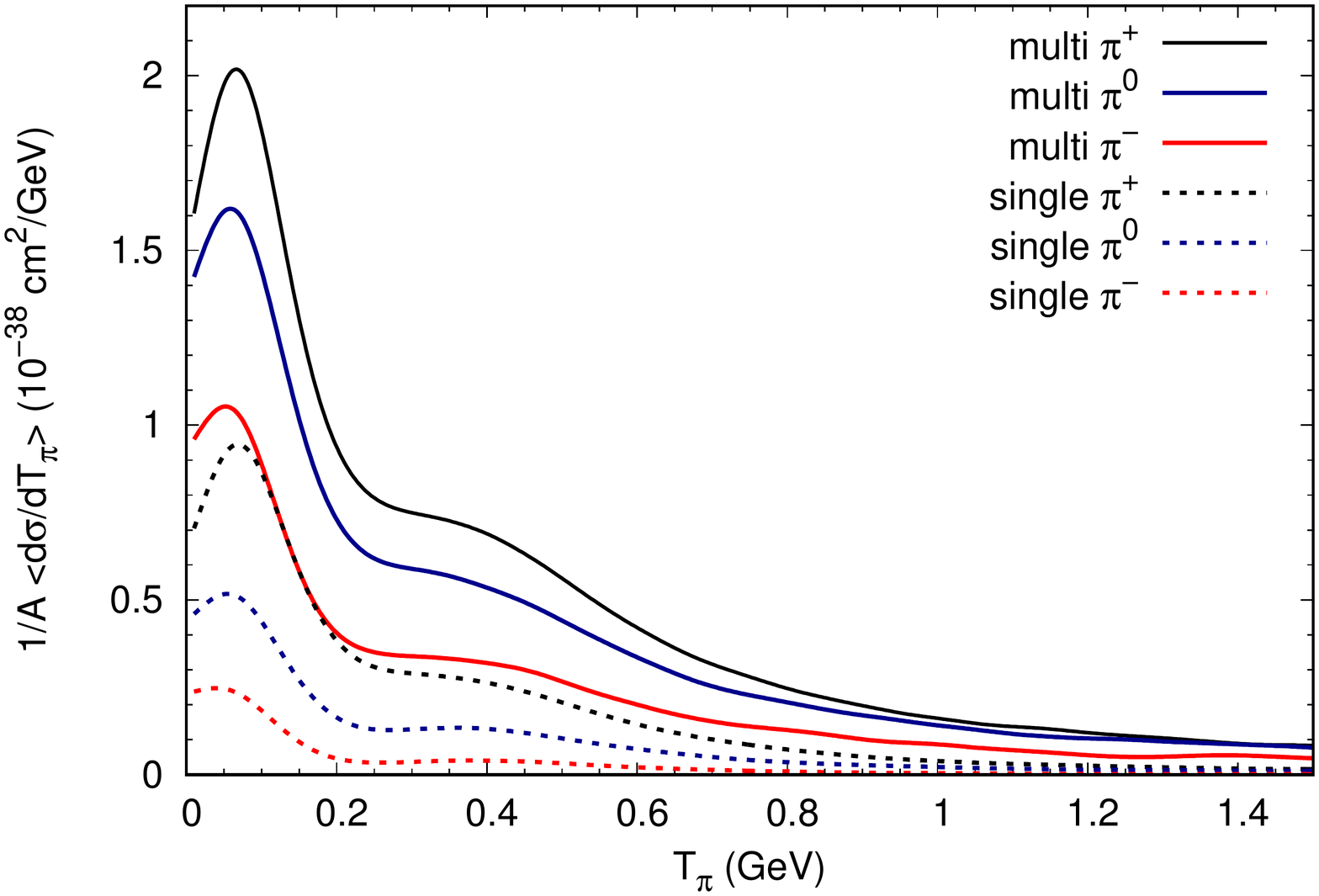}
	\caption{Kinetic energy spectra of incoherently produced pions for all charges in the DUNE ND. Shown are the cross sections both for the single and the multi-meson events.}
	\label{fig:DUNEdsigmadtpi}
\end{figure}

\begin{figure}[ht]
	\centering
	\includegraphics[width=0.8\linewidth]{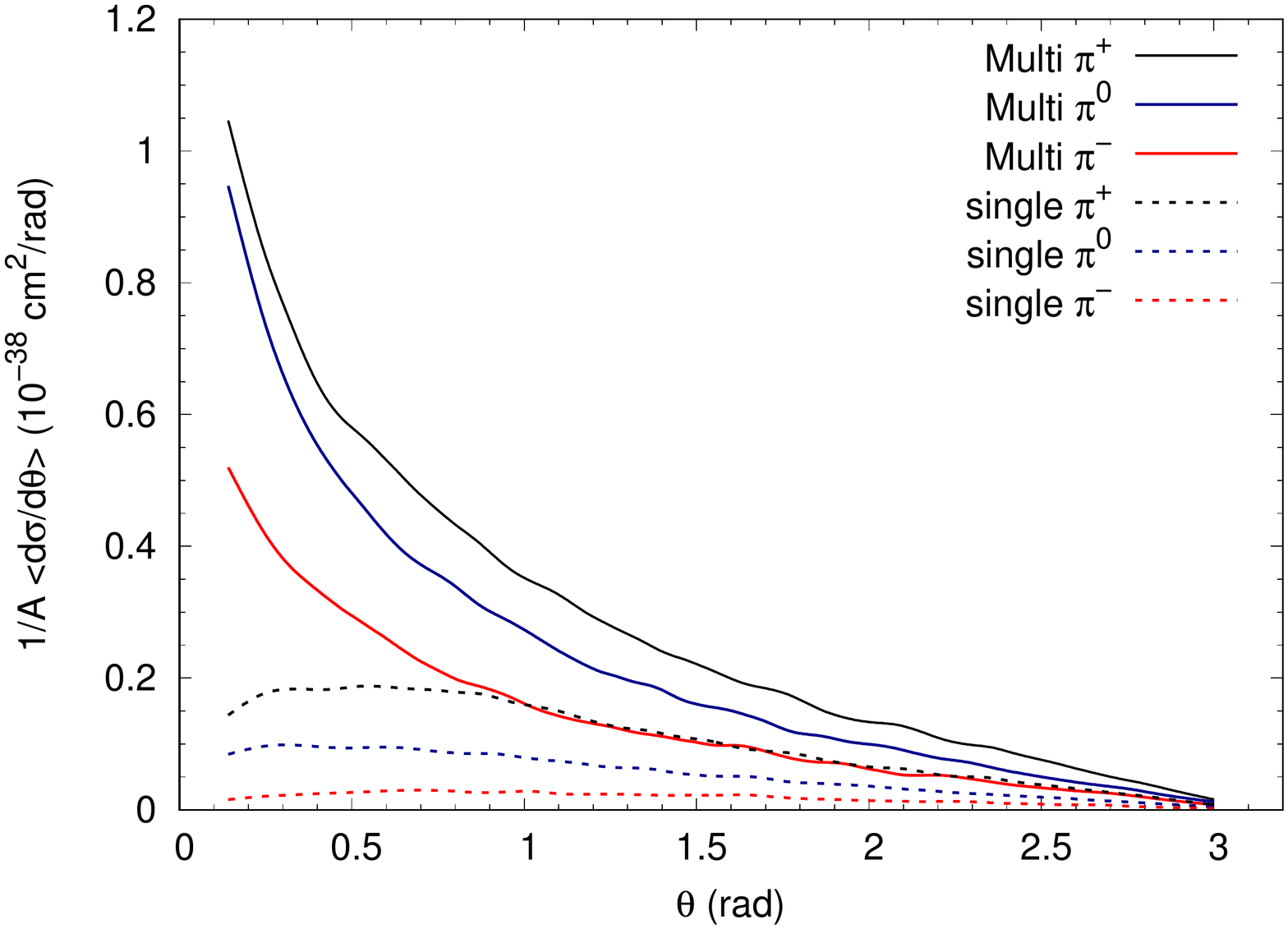}
	\caption{Angular distribution of incoherently produced pions of all charges both for single- and for multi-meson events in the DUNE ND.}
	\label{fig:DUNEsigmadthetapi}
\end{figure}

\paragraph{Pions} The following two figures \ref{fig:DUNEdsigmadtpi} and \ref{fig:DUNEsigmadthetapi} give the pion kinetic energy and angular distributions.
The pion spectra show the typical behavior known from lower energies with a strong peak at about 0.1 GeV and a flattening
around 0.25 GeV before the cross section continues to fall again towards larger kinetic energies. The flat region  is due to pion reabsorption through the $\Delta$ resonance. As expected for a neutrino beam $\pi^+$ production dominates,
but $\pi^0$ is close reflecting the larger number of neutrons compared to protons in an Ar target and the presence of charge transfer reactions. Even $\pi^-$ amounts to about 1/2 of $\pi^+$, due to the strong presence of DIS. The figures also show that -- because of the strong DIS component -- the multi-pion cross sections are much larger than the single-pion ones. All cross sections are forward peaked.

Earlier theoretical studies have shown that pions with kinetic energies below about 0.03 GeV cannot reliably be described by semiclassical methods, but require a quantum-mechanical treatment \cite{Buss:2006vh}. The DUNE experiment was originally assumed to have a threshold of about 100 MeV kinetic energy \cite{Acciarri:2015uup}. This is just about where the peak of the cross section is located. Thus, the detector cuts out a nonnegligeable part of the cross section if the cutoff cannot be lowered.

\begin{figure}[ht]
	\centering
	\includegraphics[width=0.8\linewidth]{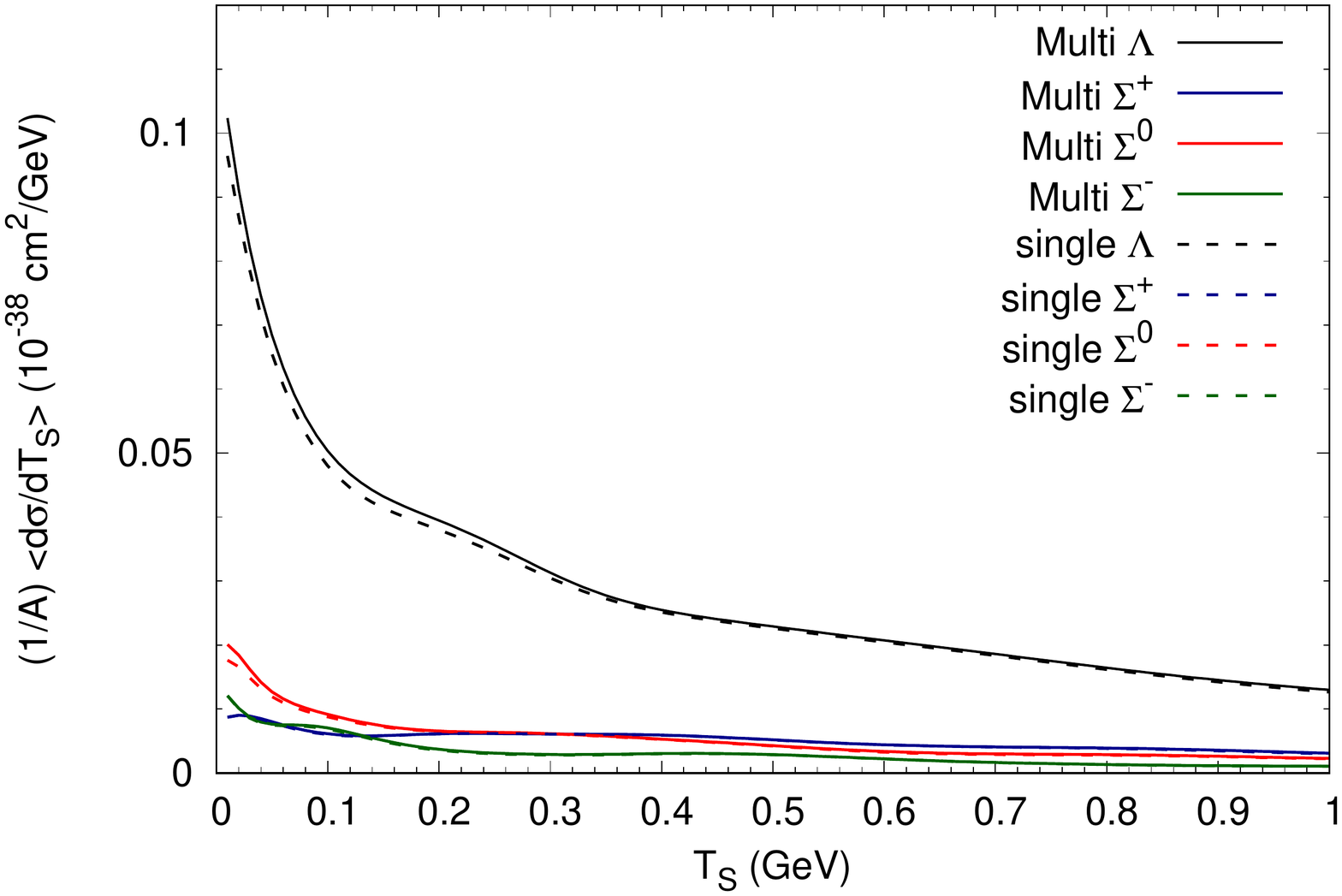}
	\caption{Spectra of strange baryons. Shown are both the spectra for multi-strange and for single-strange events in the DUNE ND. The 'single' events contain one and only one baryon of the given flavor. The 'Multi' events contain at least one strange baryon of the given flavor together with any number of other baryons.}
	\label{fig:DUNEdsigmadts}
\end{figure}

\paragraph{Strange baryons}
Fig.\ \ref{fig:DUNEdsigmadts} gives spectra for strange baryons. In this case single and multiple production cross sections lie on top of each other. This just reflects the lower probability (and higher thresholds) for strangeness production. Strangeness is produced mainly through DIS processes, often in connection with other mesons.
It is seen that $\Lambda$ production prevails. The increase of the spectrum is reminiscent of that observed above in the nucleon spectra. Indeed, the origin of that rise is again to be found in the FSI where the produced $\Lambda$ collides with the target nucleons and thereby looses energy.

\begin{figure}[ht]
	\centering
	\includegraphics[width=0.8\linewidth]{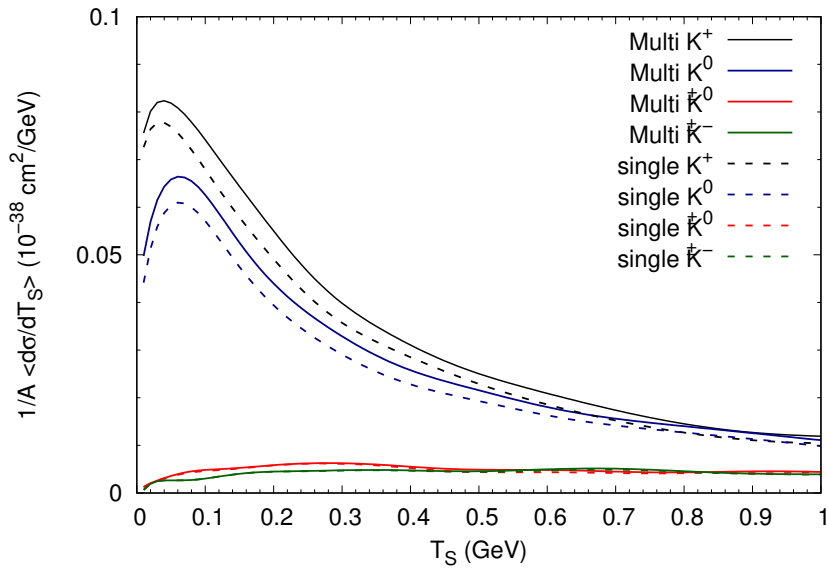}
	\caption{Spectra of strange mesons, both for single and for multi-meson events in the DUNE ND.}
	\label{fig:DUNEdsigmadtsmeson}
\end{figure}

\paragraph{Strange mesons}
The same is true for strange meson production (Fig.\ \ref{fig:DUNEdsigmadtsmeson}). There only $K^+$ and $K^0$ play any essential role; all the other flavors are much lower.

\section{Summary and Conclusion}

Generators are essential tools that make nuclear theory directly applicable to the description of neutrino-nucleus reactions. If the agreement with neutrino-nucleus reaction data is good then these generators can be relied on to do the 'backwards computation' necessary to determine the incoming neutrino energy in a broad-band oscillation experiment which in turn is needed for the extraction of neutrino mixing parameters, the mass hierarchy and a possibly CP violating phase $\delta_{CP}$ from long-baseline oscillation experiments.

In view of the importance of generators for neutrino long-baseline experiments it is surprising to see that often these generators are being used by experimenters in a 'black box mode' without much knowledge and concern about their inner workings. This information is indeed hard to come by since none of the generators GENIE, NEUT or NuWro come together with a detailed and comprehensive documentation about their physics contents and their algorithms used \footnote{For GENIE even the latest manual of version 3 \cite{GENIE:2018} often contains only headers, followed by empty pages.}. For cases where there is some text, e.g.\ for the treatment of FSI, the discussion is superficial. The reason is probably that the generators often incorporate quite old code fragments and methods for which the basic knowledge and the original authors are no longer available. This is different for GiBUU which represents a consistent framework of theory and code, is well documented \cite{Buss:2011mx} and most of its primary authors are still active in the field.

For a test of generators besides neutrino data themselves also data from electro-nuclear or photo-nuclear reactions are useful and necessary. Their initial reactions are identical with the vector part of neutrino-induced reactions and the final state interactions are the same if the incoming kinematics (energy- and momentum-transfer) are identical. Checking generators against electron or photon data is thus an indispensable requirement.

So far, only a few generators can actually be used to describe electro- or photo-nuclear interactions. There are no published results from NEUT or NuWro available for such reactions and the results obtained with GENIE are quite unsatisfactory for energy transfers beyond the QE peak \cite{Ankowski:2019}. Checks of semiinclusive cross sections with GiBUU can be found in Refs.\ \cite{Gallmeister:2016dnq,Mosel:2017ssx,Mosel:2018qmv} where they can be seen to work quite well. Agreement with such data is a necessary prerequisite for any generator, but it is not sufficient since all these electron data are semi-inclusive ones. In addition, electro-nuclear pion \cite{Kaskulov:2008ej} and $\rho$ \cite{Gallmeister:2010wn} production data must be used for a further check, as well as the many data from photonuclear pion production on nuclei \cite{Krusche:2004zc,Krusche:2004uw,Mertens:2008np,Nanova:2010sy}.

In 2017 a group of authors interested in the interplay of experimental and theoretical neutrino-nucleus physics published the following list of 'general challenges facing the community' \cite{Alvarez-Ruso:2017oui}:
\begin{enumerate}
\item The development of a unified model of nuclear structure giving the initial kinematics
and dynamics of nucleons bound in the nucleus.
\item Modeling neutrino-bound-nucleon cross sections not only at the lepton semi-inclusive
cross section level, but also in the full phase space for all the exclusive channels that are
kinematically allowed.
\item Improving our understanding of the role played by nucleon-nucleon correlations in interactions
and implementing this understanding in MC generators, in order to avoid double
counting.
\item Improving models of final state interactions, which may call for further experimental
input from other communities such as pion-nucleus scattering.
\item  Expressing these improvements of the nuclear model in terms that can be successfully
incorporated in the simulation of neutrino events by neutrino event generators.
\end{enumerate}
All of these points are indeed quite relevant.They reflect shortcomings of the  generators widely used by most neutrino physics experimenters. 

They do not, however, reflect open physics problems nor do they reflect the absence of a practical solution. Contrary to the impression one could have gotten from the discussions in Ref.\ \cite{Alvarez-Ruso:2017oui} these challenges have actually been tackled by nuclear theorists and solved about 20 years ago. The solutions have found their way into the practical implementation in the generator GiBUU.

I therefore now add some comments to the points above:
\begin{enumerate}
\item In contrast to all the other presently used generators GiBUU does have the target nucleons bound in the nucleus. In its present version the phase-space distribution of these particles is semi-classical (local Thomas-Fermi gas in a mean field potential), but this initial state could be replaced by any more refined model, e.g.\ from nuclear many-body theory. The agreement with data reached already with the present model indicates that presently available neutrino data are not very sensitive to details of the initial phase-space distributions and spectral functions. This is so because, on one hand, the smearing over incoming energies, necessarily present in all neutrino experiments, smears out all quantum-mechanical phases. On the other hand, the presently available neutrino data still carry large uncertainties. Choosing particularly sensitive observables may also change the situation in the future.
\item Generators have to provide this information on the final state and indeed all the widely used generators do that, at the expense, however, of patching up one model for the initial interactions with another, different model for the final state interactions. Models built for lepton semi-inclusive cross sections, such as the GFMC calculations and the SUSA model calculations, cannot give that information.
\item There is so far no indication for the presence of nucleon-nucleon short range correlations in any neutrino data. In electron experiments such correlations show up in quite exclusive reactions that fix energy and momentum transfer in reactions restricted to one or two outgoing nucleons. Such exclusivity cannot be achieved in neutrino experiments, simply because of the broad energy distribution in the incoming beam which naturally also leads to a broad smearing over energy transfers. Observables that are reasonably insensitive to the incoming energy may offer a way to more exclusivity.
\item Final state interactions indeed have to be checked against data obtained in other experiments. Pion-nucleus data have been used for generator checks; equally essential are particle production data from pA experiments. As discussed earlier, however, more stringent would be checks against photo- and electro-nuclear data since photons populate the whole nucleus whereas incoming pions suffer strong initial state interactions.
\item This is not a future program: all of these points are implemented in GiBUU.
\end{enumerate}

\subsection{Future developments}
The ongoing and planned long-baseline neutrino experiments all strive for a precision-dominated determination of neutrino mixing parameters. These parameter can be obtained from the experimental observations only by means of a generator which has to be as up-to-date as the experimental hardware is. One cannot stress this point strongly enough: without a reliable state-of-the-art generator the experiments cannot reach their goals. It is thus time to combine scientific expertise from nuclear theory with resources mainly from the high-energy experimental community to construct a new generator which could be built on the experience reached with GiBUU, but also some of the QGP generators.

This new generator has to fulfill the following physics requirements
\begin{itemize}
\item Foremost, it has to be built on consistent nuclear theory, with one and the same ground state for all reaction processes. This requirement removes artificial and unphysical tuning degrees of freedom.

\item It has to include potentials, both nuclear and Coulomb, from the outset. The most obvious feature of nuclei is that they are bound; generators should respect that in the preparation of the ground state. The potentials are well determined from other nuclear physics experiments; there is then no more freedom to introduce artificial binding energy parameters. Potentials will necessarily increase the computing times; this is the (small) price one has to pay for a realistic nuclear physics scenario.

\item The starting point for any such generators should be transport theory. Transport theory is no longer some esoteric theoretical 'dream', but it is well established in other fields of physics as well as in nuclear physics where all the top-level experiments searching for the quark-gluon plasma (QGP) at RHIC and LHC use generators built on transport theory.

\item If sophisticated spectral functions of bound nucleons actually become essential for the description of neutrino-nucleus reactions then modern transport theory provides the only consistent method to perform the off-shell transport necessary to bring nucleons back to their mass-shell once they leave the nucleus. The implementation of off-shell transport in actual codes is one of the major achievements of transport theory during the last 20 years \cite{Leupold:1999ga,Buss:2011mx,Cassing:2008nn}. It has found its way into the transport codes used by the QGP community.

\end{itemize}

From a more practical point of view any new generator should have the following properties in addition

\begin{itemize}

\item The new generator must be well-documented, both in its physics content and its algorithms. In addition to a well-structured complete documentation a thorough comment structure in the code is necessary. After all, the neutrino generator would have to be used over the next 20 - 30 years, most probably also by physicists who were not involved in the writing of the original code. Maintenance of the code is then only possible, if documentation and comment structure are available.

\item The new generator should allow for easy variations of essential parameters. Elementary cross sections are an essential input into any generator and these cross sections carry experimental uncertainties. It, therefore, must be possible to vary them within their uncertainties to see the effect on the final oscillation parameters. To be distinguished from that must be the tuning of unphysical or redundant parameters which can, for example, appear when different subprocesses are described by very different theories and general principles such as time-reversal invariance are not respected.

\item  The code also must have a modular structure, with well defined interfaces, that allows easier modification if new theories or algorithms for indivicual subprocesses become available. Special care must be taken, however, that the individual modules respect the intrinsic connections between different processes, such as pion production and absorption. 
    
There has to be some scientific supervisory structure that guarantees that the various processes are described on a consistent basis. A 'platform model' in which various different groups provide alternative descriptions of individual processes may work for the technical problems such as determining detector efficiencies. It cannot work for a theoretical understanding of data and for the ultimate task of extracting the relevant neutrino properties from long-baseline experiments.

\item Existing generators contain important parts which deal with the complications caused by the broad neutrino beam and the extended target sizes; both are features not present in nuclear physics experiments and outside of the expertise of nuclear theorists. These parts of the codes are essential and should be taken over into any new generator.

\end{itemize}

Any such new generator has to be checked against data not only from electro-nuclear experiments, but also from neutrino-nucleus interaction experiments. Since generators are used in the experiment for a determination of detector efficiencies there is the danger of a 'logical loop' in which a generator is first used to get the data points and then is used again to compare the data with. This is now common practice in neutrino physics experimental groups.

Instead, a 'nuclear physics model' should be implemented in which programs such as GEANT are used to determine purely experimental properties, such as efficiencies and threshold, but the final comparison of data takes place with 'real theory'.

Any future, newly built neutrino generator will have to be tested against data obtained nowadays in experiments such as MINERvA or the ND experiments in T2K or NOvA. This test will be difficult if the published data depend on generator {\it X} version {\it y.z} for which no documentation exists. Care must, therefore, be taken that the data contain as little 'generator contamination' as possible. This is, for example, not the case if cuts on incoming neutrino energies or on invariant masses are imposed on the published data since such cuts can only be imposed by means of a generator.

\ack
I gratefully acknowledge many extremely helpful and productive discussions with Kai Gallmeister, both about the physics and the inner workings of GiBUU. I am also indebted to some of my experimental colleagues for explaining to me details of data and experiments. Here,  Debbie Harris, Xianguo Lu and Kevin McFarland have been invaluable discussion partners. I am also grateful to Luis Alvarez-Ruso for a careful
reading of the manuscript.

\section*{References}

\bibliography{nuclear}

\end{document}